\documentclass[a4paper,11pt]{article}
\pdfoutput=1 

\usepackage{jheppub} 

\usepackage[T1]{fontenc} 

\newcommand{\be}{\begin{equation}}
\newcommand{\ee}{\end{equation}}

\newcommand{\mt}[1]{\textrm{\tiny #1}}

\def\pt{{p_\mt{T}}}
\def\ponet{{p_\mt{T1}}}
\def\ptwot{{p_\mt{T2}}}

\def\AJ{{A_\mt{J}}}

\def\Raa{R_{\mt{AA}}}
\def\Rcp{R_{\mt{CP}}}

\def\aC{{{\kappa}_{\rm coll}}}

\def\aR{{\kappa_{\rm rad}}}

\def\aSC{{\kappa_{\rm sc}}}

\title{\boldmath A Hybrid Strong/Weak Coupling Approach to Jet Quenching}

\author[a]{Jorge Casalderrey-Solana,}
\author[b]{Doga Can Gulhan,}
\author[c,d]{Jos\'e Guilherme Milhano,}
\author[a]{Daniel Pablos,}
\author[b,e]{Krishna Rajagopal,}

\affiliation[a]{
Departament d'Estructura i Constituents
de la Mat\`eria and Institut de Ci\`encies del Cosmos (ICCUB),
Universitat de Barcelona, Mart\'\i \ i Franqu\`es 1, 08028 Barcelona, Spain
}

\affiliation[b]{Laboratory for Nuclear Science and Department of Physics, Massachusetts Institute of Technology (MIT), Cambridge, MA 02139 USA}

\affiliation[c]{CENTRA, Instituto Superior T\'ecnico, Universidade de Lisboa, Av. Rovisco Pais, P-1049-001 Lisboa, Portugal}
\affiliation[d]{Physics Department, Theory Unit, CERN, CH-1211 Gen\`eve 23, Switzerland}

\affiliation[e]{
Center for Theoretical Physics, MIT, Cambridge, MA 02139, USA
}
\emailAdd{jorge.casalderrey@ub.edu}

\emailAdd{dgulhan@mit.edu}

\emailAdd{krishna@mit.edu}

\emailAdd{guilherme.milhano@tecnico.ulisboa.pt}

\emailAdd{dpablos@ecm.ub.es}

\preprint{{\footnotesize MIT-CTP-4550, CERN-PH-TH-2014-089, ICCUB-14-051}}

\abstract{
We propose and explore a new hybrid approach to jet quenching in a strongly coupled medium. The basis of this phenomenological approach is to treat physics processes at different energy scales differently. The high-$Q^2$ processes associated with the QCD evolution of the jet from
its production as a single hard parton through its fragmentation,
up to but not including hadronization, 
are treated perturbatively following DGLAP evolution, to which we ascribe a spacetime structure. The interactions between the partons in the shower and the 
deconfined matter within which they find themselves 
lead to energy loss. 
The momentum scales associated with the medium itself (of the order of the temperature)
and with typical interactions between partons in the shower and the medium are sufficiently soft  that 
strongly coupled physics plays  an important role in energy loss.
We model these interactions 
using qualitative insights inferred from holographic calculations of the energy loss of energetic light quarks and gluons
in a strongly coupled plasma, obtained via gauge/gravity duality.
We embed this hybrid model into  a hydrodynamic
description of the spacetime evolution of the hot
 QCD matter produced in heavy ion collisions and confront its predictions with experimental results for 
 a number of observables that have been measured in high energy jet data from heavy ion collisions at the LHC,
 including jet $\Raa$ as a function of transverse momentum, the dijet asymmetry, and the jet fragmentation function ratio, all as functions
 of collision centrality.   The holographic
 expression for the energy loss of a light quark or gluon that we incorporate in our hybrid model
 is parametrized by a stopping distance.  We find very good agreement with all the
 data as long as we choose
 a stopping distance that is comparable to but somewhat 
 longer than that in ${\cal N}=4$ supersymmetric Yang-Mills theory.
For comparison, we also construct analogous alternative models in which we assume that energy loss occurs
as it would if the plasma were weakly coupled. We close with suggestions of observables that could
provide more incisive evidence for, or against, the importance of strongly coupled physics in jet quenching.
}

\begin{document} 
\maketitle
\flushbottom

\section{Introduction}
\label{sec:intro}

One of the most striking results obtained from heavy ion collisions at the Large Hadron Collider  (LHC) is the strong suppression of high energy jets observed in Pb-Pb collisions with a center
of mass energy of 2.76 TeV per nucleon-nucleon collision~\cite{Aad:2010bu, Chatrchyan:2011sx}. This suppression, commonly referred to as jet quenching, is due to  the energy loss suffered by the components of the jets on their way out of the hot QCD medium formed in a high energy heavy ion collision. The phenomenon of jet quenching was discovered prior to the LHC measurements, without reconstructing individual jets, 
primarily via the strong reduction in the number of intermediate-$p_T$ hadrons in heavy ion collisions at RHIC relative to proton-proton collisions~\cite{Adcox:2001jp,Adler:2002xw}. Jet quenching has come to be  seen as one of the most powerful experimentally accessible tools with which to analyze the properties of deconfined QCD matter. The large magnitude of the effects of energy loss observed in heavy ion collisions at the LHC, together with the ability to study the effects of energy loss on many properties of individually reconstructed jets,
increases the potential of these probes to provide accurate medium diagnostics, provided the mechanism by which they interact with the medium can be understood with sufficient precision.

One of the reasons why high energy jets are superior to other probes is that their production occurs at very high energy scales,  $Q \gg \Lambda_{QCD}$, which guarantees that their  production spectrum is under good theoretical control, since it can be determined via perturbative QCD.  Similarly, many of the properties of jets in vacuum are also controlled by physics at high energy scales and are therefore 
well understood theoretically.  Therefore, observed deviations of those properties in a heavy ion environment must be due to the interaction of the different jet components with the hot hadronic medium that the nascent 
jet traverses on its way out of the collision zone. In general, the interaction with the medium constituents 
will lead to the degradation of the jet energy, but the precise mechanism or mechanisms 
by which this occurs depend on the nature of the medium.

Although the production of a hard parton that will become a jet, and the fragmentation
of that parton as it propagates, are controlled by weakly coupled physics at high
momentum scales, the physics of the medium produced in experimentally realizable 
heavy ion collisions is not weakly coupled.  At sufficiently high temperatures
the quark-gluon plasma must be a weakly coupled plasma of quark and gluon
quasiparticles. However, in the temperature range explored by current colliders, namely $T\sim 150-600$~MeV,
we know from the comparison of more and more precisely measured experimental observables to more and more sophisticated
calculations of relativistic viscous hydrodynamics that the quark-gluon plasma produced 
in heavy ion collisions is a droplet of strongly coupled liquid that expands and flows
collectively, hydrodynamically (For a review, see Ref.~\cite{Heinz:2013th}).
This fact makes the quark gluon plasma a very interesting form of matter that has attracted the interest
of scientists in other fields in which other forms of strongly coupled matter arise. However, this
fact also complicates the theoretical understanding of the properties and dynamics of the medium
rather significantly.
For this reason, in recent years there has been a growing interest in strongly coupled techniques 
that can shed light on the dynamics of the liquid plasmas that arise as the 
hot deconfined phases of other non-Abelian gauge theories which have holographically dual
descriptions as gravitational theories in $4+1$-dimensional spacetimes containing a black
hole horizon.  The simplest example to which this gauge/gravity duality has
been applied is the plasma that arises at nonzero temperature in strongly coupled
$\mathcal{N}=4$ supersymmetric Yang Mills  (SYM) theory in the limit of a large number
of colors $N_c$. Holographic analyses performed in this and other gauge theories
have led to many qualitative insights into the properties of the QCD plasma, its
dynamics in heavy ion collisions, and the dynamics of probe particles propagating
through the strongly coupled plasma.  (See Ref.~\cite{CasalderreySolana:2011us} for a review). 

The way in which a high energy excitation interacts with a deconfined non-Abelian plasma is well understood in two 
extreme, and unrealizable, limits. At weak coupling, by which we mean at unrealizably high temperatures
at which the coupling constant at the
medium scale is small, perturbative analyses show reliably 
that the dominant mechanism of in-medium energy loss is the radiative process of stimulated gluon emission caused 
by the scattering of the high energy parton 
off particles in the medium~\cite{Baier:1996kr,Baier:1998kq,Gyulassy:2000er,Wiedemann:2000za,Wang:2001ifa,Arnold:2002ja}. The rate of emission of these radiated gluons forms the basis of most current analysis of jet modification in the environment produced in heavy ion collisions. (See Refs.~\cite{Mehtar-Tani:2013pia,Majumder:2010qh,CasalderreySolana:2007zz,Jacobs:2004qv} for  reviews.) 
In addition, many of these studies also include a second energy loss process, that is 
in principle subleading for very high energy partons, namely
the elastic transfer of energy to medium constituents, referred to as collisional energy loss~\cite{Wicks:2005gt}.  
The second unrealizable limit is the limit in which the coupling constant is assumed to
be large at all relevant energy scales.  
In this case,
gauge/gravity duality has made it possible to use holographic calculations
to analyze the way in which varied energetic probes have their energy
degraded, and are otherwise modified, as they
propagate through strongly coupled 
plasma~\cite{Herzog:2006gh,CasalderreySolana:2006rq,Gubser:2006bz,Liu:2006nn,Chesler:2008uy,Gubser:2008as,Arnold:2010ir,Arnold:2011qi,Chesler:2011nc,Chernicoff:2011xv}. 
(For a review, see Ref.~\cite{CasalderreySolana:2011us}.)
These
computations provide detailed dynamical information on the energy loss processes in this limit.  
The intuition that comes from these calculations is phrased
in terms of the dual gravitational description, rather than in terms of gauge theory degrees of freedom.
While these two extremes each provide invaluable guidance to understanding energy loss processes in a heavy ion 
environment, because the medium is strongly coupled while much of the physics of jets is governed by weakly coupled high momentum physics, at least as they are currently constituted 
neither approach can capture all important aspects of the dynamics.

The main difficulty in understanding jet dynamics in a strongly coupled QCD medium resides in the interplay between physics at  very different energy scales. After their production via a (very) hard scattering, 
jets relax their large initial virtuality down toward the hadronic scale via an evolution process of
branching into a shower of partons.  In vacuum, this fragmentation process
is governed by the Dokshitzer-Gribov-Lipatov-Altarelli-Parisi (DGLAP) equation. This perturbative process is crucial to understanding most 
jet properties. In the medium, this evolution occurs while at  the same time partons in the developing shower suffer many soft exchanges of momenta of order the medium temperature $T$, which alter the fragmentation pattern. 
Since the momenta transferred in these interactions are not large, this physics is not weakly coupled
just as the physics of the medium itself is not weakly coupled.
This means that a part of the dynamics of jets propagating through the medium produced
in a heavy ion collision is out of the regime of validity of perturbative QCD.
Thus, jets are multi-scale probes sensitive to both strongly and weakly coupled physics.
In the long run, their description in controlled calculations
will require either a strongly coupled approach to
far-from-equilibrium dynamical processes in QCD or 
calculations done via gauge/string duality that incorporate
asymptotic freedom at short distance scales or both.  As, at present, neither seems on the horizon
we must limit our goals.
A successful phenomenological model that describes 
the modifications of jets in the medium, today, must 
be a hybrid model in which one can simultaneously treat the weakly coupled physics
of jet production and hard jet evolution and the strongly coupled dynamics of the
medium and the soft exchanges between the jet and the medium.
In this work, we will put forward a phenomenological approach which combines different physics mechanisms at different scales. While there have been other attempts to combine results obtained from weak and strong 
coupling~\cite{Marquet:2009eq,Betz:2011jp,Ficnar:2012yu,Betz:2014cza}, 
our approach is distinct since it focusses on 
using different calculational frameworks at the different energy scales.

This paper is organized as follows: we describe how we set up our hybrid approach  in Section~\ref{aha}. The interaction of partons with a strongly coupled medium is reviewed in Section~\ref{elm}. 
In Section~\ref{MC} 
we discuss how to implement these ideas in a simple Monte Carlo simulation of jets in  heavy ion collisions, using
a hydrodynamic description of the spacetime dynamics of the medium. 
We use this implementation of our hybrid approach 
to determine several jet observables, which we confront with data on jet $\Raa$, the
dijet asymmetry and jet fragmentation function ratios  in Section~\ref{cd}. 
In Section~\ref{sec:discussion} we reflect upon the successes and limitations of our
hybrid approach and, in addition, suggest further observables that, if measured, could provide more incisive
evidence for or against the importance of strongly coupled physics in jet quenching.

\section{\label{aha}A Hybrid Approach to Jet Quenching}

As we have stressed in the preceding Introduction, no single theoretical framework is currently available within which controlled calculations of all important aspects of jet quenching in heavy ion collisions can reliably be carried out.
This is so since we must simultaneously
describe the perturbative dynamics at short distances and the strongly coupled physics at the 
medium scale.
We will therefore resort to phenomenological modeling of the 
main physical processes occurring during the propagation of high energy partons through 
strongly coupled plasma. To simplify our analysis, we will focus on high energy, high virtuality jets, 
since a large separation between the hard and 
medium scales allows us to better separate the treatment of these two regimes.  In this Section, 
we will spell out and motivate the main assumptions behind our model.

Our first assumption is that the exchange of momentum with the medium, which in the absence of coherence effects among several plasma constituents is of the order of the temperature $T$, is smaller than 
the virtuality of any of the  jet partons at any stage of the evolution. 
For sufficiently high energy jets, this assumption is certainly valid at the early stages of the evolution process, 
but it becomes more questionable at the late stages, when the evolution approaches the hadronization scale. 
Fortunately, these late stages also happen at later
times, when almost all the partons in the shower are outside of the medium~\cite{CasalderreySolana:2011gx}. 
Since these small momentum exchanges cannot lead to a significant variation of a parton's 
virtuality, we will assume that the splitting kernel at each point 
in the evolution is as in the vacuum.  This motivates our second assumption: 
because each
splitting that occurs as the original parton fragments 
happens at smaller distance scales than the medium can resolve,
we assume that the splitting probabilities are as in vacuum.
Keeping the splitting kernel unmodified implies, in particular, that, in a 
probabilistic approach, the emission probability at each step in the Markovian chain  
remains independent of the medium dynamics.

It will be important to return to the second assumption above in future work for at least two
reasons.  First, we will be assuming that the splitting probability is unmodified even
as the partons lose some of their energy and will thus be neglecting the fact that
even in vacuum the splitting kernel depends on parton energy (through
Sudakov logs, which is to say via the phase space for splitting).
Second, we will be neglecting the possibility of additional splitting induced by multiple soft exchanges with the strongly coupled plasma, which accumulate into a hard momentum transfer. As such an effect is known to be important in a weakly coupled plasma with 
point-like constituents, it will, in the future, be interesting to investigate how to incorporate it  
within the hybrid model we are setting up in the present paper.
However, assuming the physics at the medium scale to be 
strongly coupled, as we shall do throughout, renders any  
such weakly coupled
large momentum transfer processes, and their modification, 
subleading in their consequences.

We now wish to apply a prescription for how much energy each parton in the 
shower loses as it propagates through the medium.  That means that we need to
know the temperature of the medium in which a particular parton in the shower
finds itself, which in turn means that we need to know where each parton in
the shower is in space and time.
The DGLAP evolution equations that describe the fragmentation of the 
parent parton and the evolution of the resulting shower are derived
in perturbative QCD in momentum space.
They contain 
little information about how the process of showering, and the attendant
relaxation in the virtuality of the individual partons in the shower, develops
in space and time.
This space-time information is unimportant in vacuum physics, since 
the partonic components of the jet do not interact with anything; all they do is fragment and in describing the jet in the final state it is completely unimportant where and when each splitting happened. 
However, in a heavy ion environment before the shower emerges
from the medium every parton in the shower interacts with the medium, and the medium
itself changes as a function of space and time.
We therefore need to know where and when each splitting occurs. 
Based on the analysis of soft gluon emission, most jet Monte Carlo studies assign a time to each 
rung  of the evolution equation  
related to the formation time of the emitted gluon $\tau_f = 2 \omega/k_\perp^2$. 
However, the detailed implementation 
varies from one Monte Carlo implementation to another, 
which gives a sense of the theoretical uncertainty concerning the space-time evolution that is
common to all in-medium event generators. 
In this work, we will use the prescription of  Ref.~\cite{CasalderreySolana:2011gx} and  
assign a life-time to each rung of the decay chain ({\it i.e.}~to each parton in the shower)
determined from their virtuality $Q$ and energy $E$ as 
\be
\label{eq:time}
\tau= 2 \frac{E}{Q^2}\,,
\ee
with the factor of two chosen such that in the soft limit it coincides with the standard expression
for the formation time.
We will also assume that the strong virtuality ordering in the QCD shower translates into time ordering, with the 
hardest splittings occurring first. This implies that the later stages of the evolution, for 
which the virtuality is closer to the hadronization scale,  occur  at later times.

In between any of the virtuality relaxing splittings, the partons in the jet propagate through
the strongly coupled plasma. The momenta exchanged between these partons and the medium is of order the medium temperature, and therefore, for plasma temperatures not far above the 
deconfining transition, these momentum exchanges are not weakly coupled processes.
This is where strongly coupled dynamics plays a role. 
From the point of view of the jet shower, the medium takes energy away from 
each of the propagating partons and rapidly turns that energy into heating of, and collective
motion of, the medium itself -- which is to say extra soft particles in the final state, moving
in random directions.  This directly yields 
a reduction in the overall energy of the jet. 
This is in stark contrast with the perturbative mechanism of radiative energy
loss, where energy is lost through medium induced radiation of gluons 
with momenta
that are well above the medium scale and that are typically almost collinear
with the initial hard parton when they are produced.  
This radiative loss of energy by
the hard parton translates into a loss of energy for the jet in the final state
only if the radiated gluons are either (atypically) produced at large angles
relative to the direction of the hard parton or if the radiated gluons are
deflected by their further interactions with the medium~\cite{CasalderreySolana:2010eh,Blaizot:2013hx}.

\begin{figure}
\centering 
\includegraphics[width=.9\textwidth]{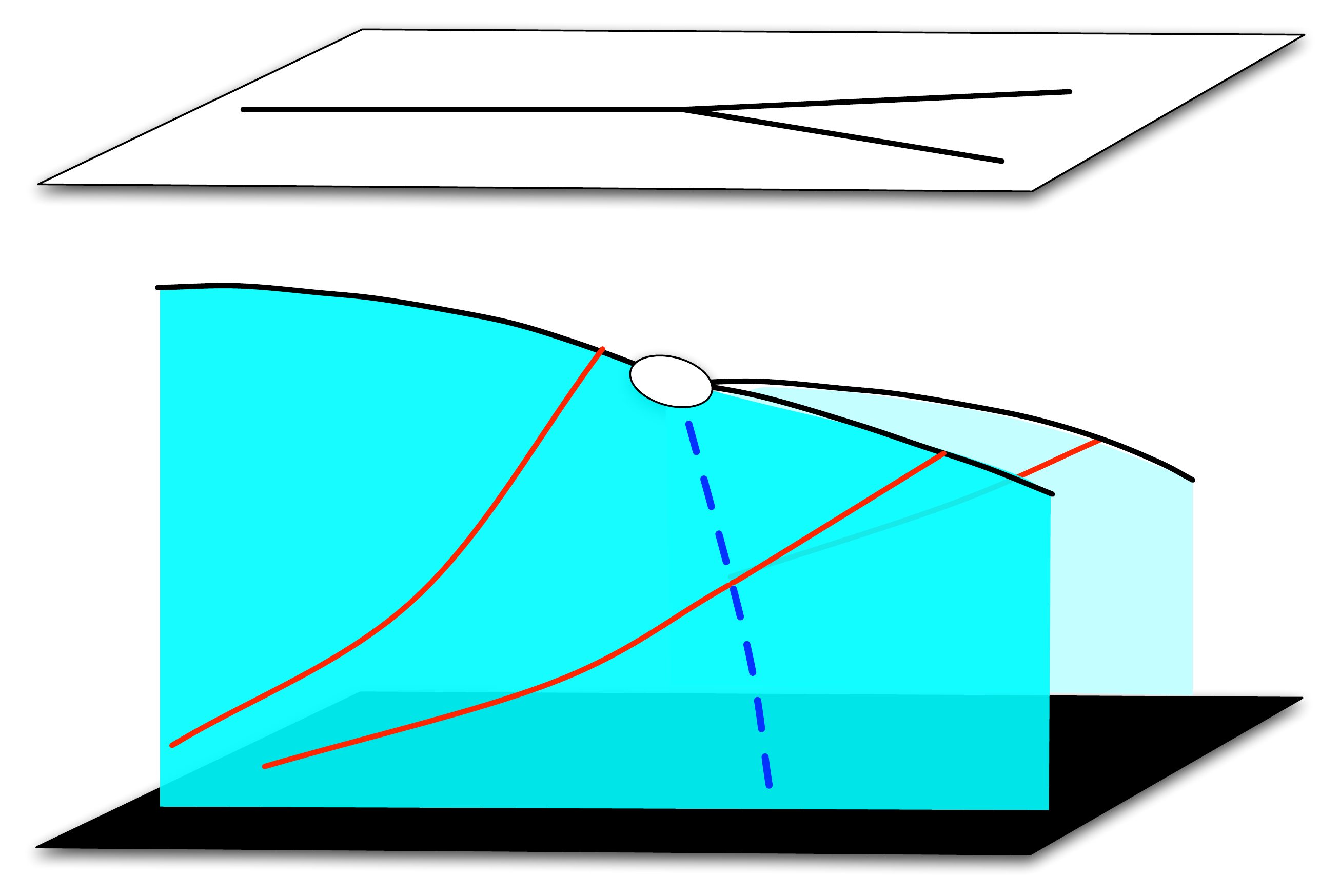}
\put(-95,255){\Large Gauge Theory}
\put(-200,235){DGLAP}
\put(-75,65){\Large Horizon}
\put(-301,107){Falling}
\put(-300,95){String}
\put(-202,171){Induced}
\put(-200,159){Vertex}
\caption{\label{Sketch}  Sketch of two views of the interaction of a high energy jet with the strongly coupled plasma. In the gauge theory, represented by the white plane at the top of the figure, 
an energetic virtual parton propagates through the medium, loses energy, and splits via (vacuum) DGLAP evolution. 
There is no (easy) way to describe the strong soft interactions between the partons and the medium in
this representation.
In the dual gravitational view, represented below, the soft interactions are represented by
a string trailing behind each parton, transporting energy from the parton ``down''
 to the horizon which is represented
by the black plane at the bottom of the figure. The parton itself, represented by the endpoints
of the string which is to say by the black lines, is also pulled ``downward'' toward the horizon.
In this representation, there is no (easy) way
to describe the splitting of one string into two, which is to say the hard splitting process
in the gauge theory. 
 In the picture we represent the splitting vertex at which one
string becomes two by a white oval below which a hypothetical string-splitting curve
shown as a blue dashed line trails.
At present there is no known calculation underlying these aspects
of the figure.
}
\end{figure}
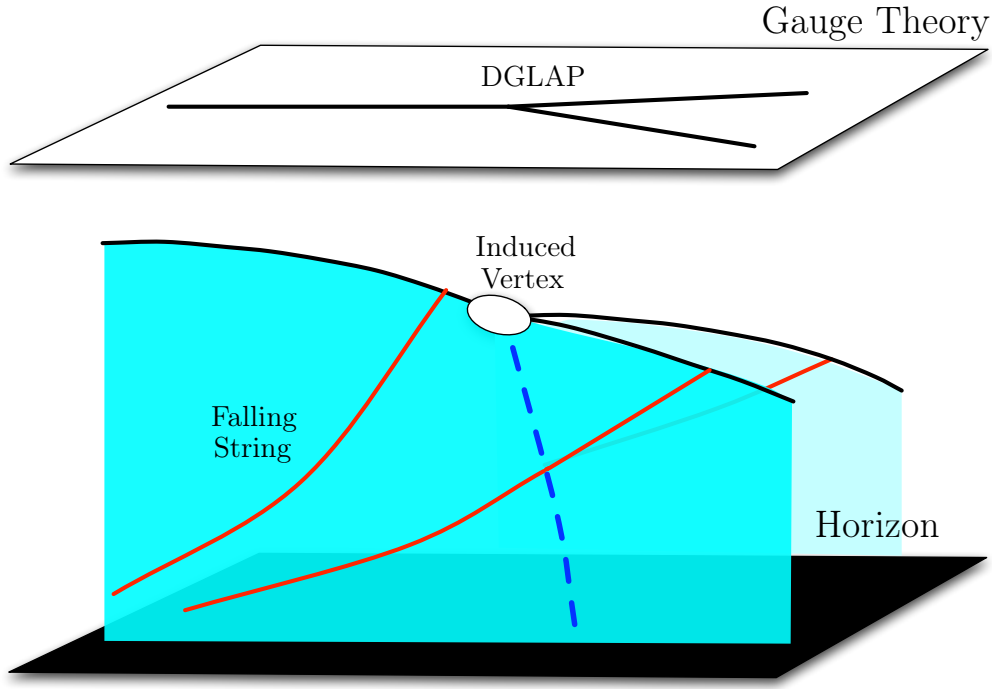

In a nutshell,  we shall assume that 
no hard radiative processes occur between the DGLAP vertices and  that the dynamics of these partons in the 
plasma is analogous to that of energetic objects propagating through the strongly coupled plasma in
a gauge theory with a dual gravitational description.
While the theories that possess a gravity dual do not yet include QCD, we will utilize the powerful ideas of the duality to gain qualitative understanding of the relevant strongly coupled dynamics, an approach that has proved
useful in many contexts. (See Ref.~\cite{CasalderreySolana:2011us} for a review.)
However, keeping in mind that these calculations are not done in QCD itself, we will
use the explicit results obtained via holographic calculations only as indicative, specifically by keeping
all of their parametric dependences while introducing one free dimensionless parameter
that we shall fit to data.  
We shall describe how we do this in concrete terms over the course of Sections~\ref{elm} and \ref{cd}.

One important aspect of how we  set up our hybrid model is minimalism. We will keep, as much as possible,  
only well-understood weakly coupled and strongly coupled physics and introduce as few as possible, in fact only one,  phenomenological
parameter that governs how we put the two together and that
needs to be fixed by fitting to data. Introducing further physics into the model 
would on the one hand
allow us to describe some of the less important physics that, as we have described, we are leaving out
but on the other hand it would introduce further parameters.  Our goal here is to construct a hybrid
model that is, in this sense, as simple as we can make it and see how well it does when confronted
with data.

The underlying picture that we are putting forward in this paper is sketched in Fig.~\ref{Sketch}. 
In the gauge theory, any of the partons of the jet which propagate in plasma may suffer a hard splitting, 
governed by the DGLAP equations. In addition to these hard splittings, these partons possess associated soft fields that interact 
strongly with the medium. 
These have a natural interpretation in a dual gravitational representation: they are strings trailing behind the 
quark, which is represented by the end point of the string. 
As noted in Ref.~\cite{Chesler:2008uy} the string end point itself does not live on the boundary of 
the spacetime, 
but ``falls'' downward, away from the boundary and toward the horizon, as the system evolves.
In the gravitational representation, the trailing strings carry energy from the quarks ``down''
toward the horizon.  This represents the process by which each quark loses energy, energy which
subsequently thermalizes, making a little more or a little hotter plasma.  
Reading the figure from left to right, one string enters from the left, with its
shape controlled by well-understood gravitational dynamics that describes
how the single quark represented by its endpoint loses energy. 
Next, a perturbative hard splitting, described in the gauge theory
by DGLAP, occurs. It is not presently known whether, and if so how,
this splitting process can be described in the gravitational
representation.
The gravitational  description must be supplemented by some induced vertex and an associated
line along which string world sheets merge, but the form of this vertex and string merging line
are not known.
Nevertheless, the gauge theory representation demands that after this splitting process,
we have two string end points below each of which a string trails.  In fact, if one of
the daughter partons is a gluon, a double string must be formed, trailing below the corresponding endpoint.
So, somehow, the single incident string worldsheet splits into the several world sheets that describe
the decay products of the DGLAP splitting.  Again, we describe this process in the gauge theory because
it is not known how to describe it gravitationally.

Since splitting processes happen at short distances, the induced vertex must occur first as close as possible to the boundary, {\it i.e.}~where the string end point splits, and only later propagate toward the horizon as represented
in Fig.~\ref{Sketch} by the blue dashed line.
This can also be interpreted as a delay in the ability of softer modes to resolve the splitting of color charges. Nevertheless, 
since the geodesic distance in AdS from the horizon to the boundary is finite, of order $1/(\pi T)$, 
after this short time the string world-sheet is fully split and 
each of the objects propagate independently through the strongly coupled 
plasma.\footnote{
Here, we are describing a delay of order $1/(\pi T)$ in the response of
the energy loss process to a sudden change in the nature of the object losing energy,
namely the splitting process.  In Ref.~\cite{Chesler:2013cqa} a similar delay time, also
of order $1/(\pi T)$, arises (and is analyzed quantitatively) in the case where the object losing energy (a heavy
quark being dragged at constant velocity) is unchanged but the temperature of the strongly coupled medium
changes suddenly.}
After the transient behavior associated with the splitting, the energy loss of each of the daughter
partons in the strongly coupled plasma is described by the dynamics of their own
trailing string --- until each of the daughters itself splits at a new hard vertex, and the process
iterates.
We are  far from providing a firm theoretical footing for the 
hybrid physical picture we have described.  Each half of the hybrid is built upon
solid ground, but different solid ground.  
In this paper, we will explore the phenomenological consequences of these ideas in 
a simplified model implementation which we hope captures the main features of some future
complete computation.

\section{In-medium Energy Loss of Energetic Particles}
\label{elm}

The principal ingredient that remains to be specified in the description of
our hybrid model is the rate of energy loss of energetic particles in the medium.
In our model we shall apply such a prescription to each of the partons in
a shower, while those partons find themselves in a medium
with local temperature $T$, with $T$ varying as a function of space and time. 
In this Section, we specify the different prescriptions for energy loss
that we have investigated by giving them for the case of a single energetic parton
propagating through a medium with constant temperature $T$.
Our principal goal is of course to investigate the validity of the hybrid strong/weak coupling
approach to jet quenching that we have described, in which the shower develops
according to a weakly coupled prescription and each parton in it loses energy
according to a strongly coupled prescription.  However, to provide benchmarks
for our computations we shall also try employing weakly coupled prescriptions for how
each parton in the shower loses energy in our formalism 
and compare results obtained in this way to the results
we obtain in our hybrid strong/weak coupling model. 
In the two subsections below we specify the details of
the strongly coupled and weakly coupled expressions for parton
energy loss that we shall employ.

\subsection{Parton energy loss at strong coupling from falling semiclassical strings}

The problem of energy loss of energetic light degrees of freedom in strongly coupled gauge theories with a gravity dual has been studied extensively.  (See Refs.~\cite{Gubser:2008as,Hatta:2008tx,Chesler:2008wd,Chesler:2008uy,Arnold:2010ir,Arnold:2011qi,Arnold:2011vv,Ficnar:2012np,Arnold:2012uc,Arnold:2012qg,Ficnar:2013wba,Ficnar:2013qxa,Chesler:2014jva} for entries into the literature.)  
These studies can be divided into two general classes: those in which a 
hard process in a strongly coupled gauge theory is studied via
the gauge/gravity correspondence, for example via analyzing the
decay of a virtual external $U(1)$ field into strongly coupled matter
within the plasma~\cite{Arnold:2010ir,Arnold:2011qi,Arnold:2011vv,Arnold:2012uc,Arnold:2012qg}; and those in which 
single energetic excitations are described as a string moving in the 
dual gravitational spacetime whose endpoint is attached to
a space-filling D7-brane and can therefore fall into the 
horizon~\cite{Gubser:2008as,Chesler:2008uy,Chesler:2014jva}. 
The former has the advantage that the set-up is fully determined within the strongly coupled theory, 
while in the latter the initial conditions that characterize 
the hard creation of these excitations need to be specified. The latter has the advantage that the string 
describes an isolated excitation whose energy can be tracked, emerging from the initial configuration.  
These two approaches lead to qualitatively similar results for certain observables, such as the parametric 
dependence of the maximal stopping distance of energetic partons, but differ quantitatively. 
While both computations are valid within the context of strongly coupled gauge theories, it is unclear 
which is a better proxy for QCD hard processes in strongly coupled medium. Since 
the string-based computations provide  the energy loss rate explicitly~\cite{Chesler:2014jva}, 
we will adopt this second approach to construct our hybrid model.

In Refs.~\cite{Chesler:2008uy,Chesler:2014jva}, a pair of high energy `quark jets' in the 
fundamental representation of the gauge group are produced moving in opposite
directions.  In Ref.~\cite{Chesler:2014jva} the setup is such that one of the `quark jets'
is incident upon a `slab' of strongly coupled plasma with temperature $T$, 
that is finite in extent with thickness $x$.  The dual gravitational description of
the `quark jet' is provided via a string
whose endpoint falls downward into the bulk, as in the left portion of the
sketch in Fig.~\ref{Sketch}.  After propagating for a distance $x$ through
the plasma the string, which is to say the quark, emerges into vacuum.
The energy $E$ of the `quark jet' that emerges from the slab of plasma, as well
as its other properties, can be compared to the initial energy $E_{\rm in}$ 
of the parton incident upon the slab and to the properties of the `jet' that would
have been obtained had their been no slab of plasma present~\cite{Chesler:2014jva}.
For our purposes, we are interested in how the energy of the `quark jet' depends
on $x$, which is to say the rate of energy loss $dE/dx$.
If the high energy `quark' is produced
next to the slab, meaning that it enters it immediately without first propagating
in vacuum, and if the thickness of the slab is large enough that initial transients
can be neglected, meaning $x \gg 1/(\pi T)$,
the rate of energy loss is independent of many details of the string
configuration and takes the form~\cite{Chesler:2014jva}
\be
\label{eq:Elrate}
\frac{1}{E_{\rm in}}\frac{dE}{dx}= -\frac{4}{\pi}\frac{ x^2}{  x^2_{\rm stop}}\frac{1}{ \sqrt{x^2_{\rm stop} -x^2}}
\ee
where $E_{\rm in}$ is the initial energy of the `quark', as it is produced and as it is incident
upon the slab of plasma and where $x_{\rm stop}$ is the stopping distance of the
`quark'.  Since $E\rightarrow 0$ 
as $x\rightarrow x_{\rm stop}$, the expression (\ref{eq:Elrate}) is only valid for $1/(\pi T) \ll x<x_{\rm stop}$. 
The parametric dependence of $x_{\rm stop}$ on $E_{\rm in}$ and $T$
was obtained previously in Refs.~\cite{Gubser:2008as,Chesler:2008uy}. For
a string whose initial state is prepared in such a way as to yield the maximal stopping
distance for a `quark' produced
with a given $E_{\rm in}$ propagating through the strongly coupled ${\cal N}=4$ SYM plasma
with temperature $T$, it
is given by
\be
\label{eq:xstop}
x_{\rm stop}= \frac{1}{ 2 \, \aSC } \, \frac{E^{1/3}_{\rm in}}{T ^{4/3}} \,,
\ee 
where we have introduced a dimensionless constant $\aSC$, the subscript 
signifying ``Strong Coupling'',  that in the calculation of Ref.~\cite{Chesler:2008uy}
is given by $\aSC=1.05\, \lambda^{1/6}$, with $\lambda$ the \'{}t~Hooft coupling. 
In the case of a slab of plasma in which $T$, and therefore $x_{\rm stop}$ is constant,
the energy loss rate (\ref{eq:Elrate}) can easily be integrated to obtain $E(x)$~\cite{Chesler:2014jva}.  
We shall
be describing the energy loss of partons in a shower that
are propagating through a medium whose temperature is changing as a function of
space and time as in a heavy ion collision; in this context what we need from
Ref.~\cite{Chesler:2014jva} is $dE/dx$, namely (\ref{eq:Elrate}).

The energy loss rate Eq.~(\ref{eq:Elrate}) has two characteristic features that distinguish 
it parametrically from analogous perturbative expressions that describe the energy loss of a single
hard parton propagating through (a slab of) weakly coupled plasma with temperature $T$,
expressions that we shall provide in the following subsection.  First,
while $x$ is not yet comparable to $x_{\rm stop}$ the rate of energy loss 
$dE/dx$ is independent of  $E_{\rm in}$ and grows 
rapidly with $x$, with a characteristic $x^2$ dependence.   Later, though,
once $x$ has become comparable to $x_{\rm stop}$ we see that $dE/dx$ 
depends in a nontrivial ({\it i.e.}~non-power-law) way 
on both $E_{\rm in}$ and $x$ and grows 
rapidly, diverging as $x\rightarrow x_{\rm stop}$ and $E\rightarrow 0$.
We note that in spite of the simple relation between $E_{\rm in}$ and the stopping distance $x_{\rm stop}$, 
the parametric dependence of the energy loss rate on the path length $x$ is 
intricate, deviating from a simple power of the length very substantially at late times.

The energy lost by the energetic parton propagating through the strongly coupled
plasma is quickly converted into hydrodynamic excitations with wave vectors $q \sim \pi T$ and
smaller.   This happens over 
a very short time $1/\Gamma_1$, with $\Gamma_1= 2\pi T \sim T/0.16$ the width of the 
lowest non-hydrodynamical quasinormal mode of the 
strongly coupled plasma, determined in the dual gravitational theory
in Ref.~\cite{Kovtun:2005ev}.
The hydrodynamic excitations are,  in turn, dissipated as heat after 
a damping time  $3 Ts/(4 q^2 \eta)$ (for sound waves)
or $Ts/(q^2 \eta)$ (for diffusive modes)~\cite{CasalderreySolana:2004qm}.
If we take the shear viscosity to entropy
density ratio to be $\eta/s\sim 2/(4\pi)$, hydrodynamic modes with $q\sim \pi T$
dissipate over a time $\sim (0.5-0.6)/T$.  Longer wavelength modes live longer.
This means that most of the `lost' energy rapidly becomes part of the plasma,
thermalizing and resulting in a little more, or a little hotter, plasma.
From an experimental point of view, the lost energy becomes extra,
soft, hadrons with momenta $\sim \pi T$ moving in random directions.
These extra hadrons will be uniformly distributed in angle, on average,
if the passage of the jet does not induce any substantial 
collective motion of the plasma.

Because we shall focus on reconstructed jet data, which is to say measurements of
the components of the jet that emerge from the plasma, we shall
make no attempt to track the lost energy in our hybrid model.
Of course, since  the `lost' energy ends up as soft hadrons going in {\it all} directions,
some of it {\it will} end up in the jet cone.  
We will make no attempt to add soft hadrons corresponding to some of the
lost energy to the jets in our model.  The reason that we make no such addition to our jets
is that 
when experimentalists reconstruct jets from data, they use some background
subtraction procedure designed to remove soft hadrons that are uncorrelated
with the jet direction, for example subtracting an $\eta \leftrightarrow -\eta$ reflection of the event from
the real event.
This means that if the `lost' energy ends up perfectly uniformly distributed
in angle,   it will be subtracted during the jet reconstruction procedure.  
If this assumption is correct, the `lost' energy does not appear
in the jets as reconstructed by the experimentalists. We therefore
make no attempt to add it to the jets we obtain from our model.
We leave to future 
work the investigation of 
fluctuations and collective flow that can in fact result in the `lost' energy 
that is deposited in the jet cone not being fully removed during
the background subtraction, meaning that some of it ends up 
being counted as a part of the jet.  The uncertainty
associated with these considerations means that when we compute 
jet fragmentation functions in Section~\ref{cd}, they may not
be reliable
for components of the jet with momenta of order 1-2~GeV.

Although the energy loss rate $dE/dx$ in (\ref{eq:Elrate}) was derived within the string-based computation
of Refs.~\cite{Chesler:2008uy,Chesler:2014jva}, the parametric dependence 
of the stopping distance for excitations with the maximum possible stopping distance
for a given energy $E_{\rm in}$ given by the expression
(\ref{eq:xstop}) 
is common to  both ways of describing high energy excitations in the plasma discussed above, 
which makes it seem a robust expectation from strong coupling computations within the gravitational 
description of large-$N_c$ $\mathcal{N}=4$ SYM theory. 
In contrast, the explicit value of the dimensionless constant $\aSC$, and even its dependence on the
 \'{}t~Hooft coupling, are not robust in the same sense.   
 There is every reason to expect that the
 numerical value of $\aSC$ will be smaller in the strongly coupled QCD plasma 
 than in the strongly coupled ${\cal N}=4$
 SYM plasma.  And, even in the latter theory, the calculations of 
 Refs.~\cite{Arnold:2010ir,Arnold:2011qi} indicate a value $\aSC$ that is ${\cal O}(1)$, {\it i.e.}~finite
 in the $\lambda\to\infty$ limit, rather than ${\cal O}(\lambda^{1/6})$ as in the
 string-based calculation of Ref.~\cite{Chesler:2008uy}.  We shall return
 to this point in Section~\ref{sec:discussion} when we discuss the implications
 of the value of $\aSC$ that we shall obtain via comparison to data in Section~\ref{cd}.

Both Eqs.~(\ref{eq:Elrate})  and~(\ref{eq:xstop}) were derived for 
energetic particles in the fundamental representation of the gauge group,
proxies for energetic quarks propagating through the strongly coupled plasma. 
However, it is impossible to model hard processes in high energy hadronic collisions
without also having the means with which to include energetic particles in 
the adjoint representation, {\it i.e.}~energetic gluons.  In our context, regardless of the identity
of the initial parton produced in a hard scattering, the shower of partons that results
and whose energy loss we shall be following necessarily includes both gluons and quarks.
Studies of high energy particles in the adjoint representation,
modeled by double strings propagating through the plasma, were initiated in Ref.~\cite{Gubser:2008as}
and have shown that these excitations also have $x_{\rm stop}\propto E_{\rm in}^{1/3}/T^{4/3}$.
Within the string-based picture, 
because the string configuration representing an energetic gluon possesses 
two strings trailing behind the `endpoint' (actually, in this case, the point where the string
folds back upon itself) it is natural to expect that
the stopping distance for a gluon is identical to that for a quark with half the energy of the 
gluon~\cite{Chesler:2008uy}.
We will further interpret this factor 2 as the large $N_c$ limit of the ratio of the
Casimirs of the adjoint and the fundamental representations of the color gauge group. 
Given these considerations,  
we will assume that an energetic gluon has the same energy loss rate Eq.~(\ref{eq:Elrate}) 
but with the prefactor in the stopping distance (\ref{eq:xstop}) given by 
\be
\aSC^G = \aSC \left(\frac{C_A}{C_F}\right)^{1/3}
\ee
with $C_A/C_F=9/4$ the ratio of Casimirs, meaning that $x_{\rm stop}$ for gluons
is shorter than that for quarks with the same energy, but only by a factor of $(9/4)^{1/3}$.

Because of the small $1/3$ power, the difference between the rate of energy loss
of quarks and gluons is small, much smaller in the strongly coupled
plasma than would be the case in a weakly coupled plasma.
We will elaborate on the consequences of this observation in Section~\ref{cfcadep}.

\subsection{Comparison with other approaches}

The realization that the physics at the medium scale is not weakly coupled has 
motivated several previous phenomenological attempts to implement strongly coupled 
computations of the in-medium interaction of high energy particles in 
the modeling of hard processes in heavy ion collisions. 
Before we continue, it is important to compare and contrast our implementation
to those in previous work.

Some early explorations 
were based on the straightforward use of 
energy loss rates based upon results derived for a single heavy or light quark
traversing the strongly coupled plasma of a gauge theory with a holographic
description~\cite{Horowitz:2008ig,Horowitz:2011gd,Betz:2011jp,Horowitz:2011cv,Ficnar:2012yu,Ficnar:2013qxa,Betz:2014cza}. 
These computations are all aimed at describing the suppression of the production
of a single high-$\pt$ hadron, {\it i.e.} the leading hadron in a jet.  None of these
early explorations included the calculation of jet observables; we shall analyze
three complementary classes of jet observables in Section~\ref{cd}.
These early explorations also do not
include the perturbative QCD evolution of the hard virtual parton.  And, 
as they describe single partons, they cannot
address the question of how the propagation through the strongly coupled plasma
does or does not modify the jet fragmentation function, a question that we shall
find plays a significant role in differentiating between energy loss mechanisms.
Furthermore, in some cases~\cite{Horowitz:2011gd,Betz:2011jp,Betz:2014cza} the
rate of energy loss of a hard parton is assumed to be
a power law in the parton energy and the propagation
distance, whereas we now know from 
Ref.~\cite{Chesler:2014jva}
that this is true only for partons which do not travel a significant fraction
of their stopping distance, as for those and only those partons $dE/dx \propto E_{\rm in}^0 x^2 $.
The complete dependence of $dE/dx$ in (\ref{eq:Elrate})
on $x$ and $E_{\rm in}$ is very different from a power law.
In other cases~\cite{Ficnar:2012yu}, 
the energy loss rate employed was based on approximations 
to the numerical analysis of Ref.~\cite{Ficnar:2012np}, which 
do not coincide in any limit with the expression derived in Ref.~\cite{Chesler:2014jva}. 
The energy loss expressions obtained more recently in Ref.~\cite{Ficnar:2013qxa}
are complementary, in that they are derived in the dual gravitational
theory using semiclassical strings that do not satisfy standard open string boundary
conditions, meaning that it remains to be determined how they can be used
in the description of light quark energy loss.

Among the work that comes before ours, the study that is in many respects
most similar to ours is that
described in Ref.~\cite{Marquet:2009eq}, although like in the previous work
above this study focusses on hadronic observables rather than computing jet observables as we do. 
Unlike in the previous work above, this study 
involves a Monte Carlo implementation of a shower
in which partons produced at high virtualities evolve down to a hadronic scale.
However, the implementation of the strongly coupled dynamics used in Ref.~\cite{Marquet:2009eq}
is very different than our own, 
as it  is based on an early interpretation of strongly coupled energy loss in partonic-like
terms
advocated in Refs.~\cite{Dominguez:2008vd,Hatta:2008tx}.  In this 
approach, the energy loss of a hard parton in strongly coupled plasma is interpreted
in the language of radiative energy loss, except with a momentum transfer from the
plasma which grows linearly with propagation distance. (In the standard weakly
coupled perturbative analysis of radiative energy loss, it is the square of the momentum
transfer which grows linearly with propagation distance.)
Based upon this earlier work,
the authors of Ref.~\cite{Marquet:2009eq} assumed an energy loss
mechanism in which weakly coupled high momentum gluons are radiated (as
at weak coupling) but in which the momentum transverse to the jet direction
that is transferred to the radiated gluons accumulates linearly with
propagation distance (unlike at weak coupling).
So, although we follow Ref.~\cite{Marquet:2009eq} in the sense that we are developing
a hybrid model that melds together features of energy loss in a strongly coupled gauge theory
with a Monte Carlo (in our case PYTHIA) implementation of perturbative splitting  in a parton shower,
our implementation of the strongly coupled physics is completely different than that in Ref.~\cite{Marquet:2009eq},
since we (i) treat all strongly coupled processes as occurring at soft, nonperturbative, scales;
(ii) use the energy loss rate derived from a complete strong coupling computation that
was not yet available at the time of the study in Ref.~\cite{Marquet:2009eq}; and
(iii) incorporate a feature that is by now understood to be characteristic
of energy loss in a strongly coupled plasma, namely 
that the `lost' energy becomes extra heat
or extra plasma, which is to say soft particles whose
directions are uncorrelated with the jet direction.

\subsection{Perturbative benchmarks: radiative and collisional energy loss in a weakly coupled plasma}

To gauge the sensitivity of the classes of jet measurements that we will use to constrain our hybrid approach, 
we wish to compare its results to those in which we replace the strongly coupled result (\ref{eq:Elrate})
for the energy loss rate of a parton in the shower with a perturbatively inspired expression for $dE/dx$.
We shall in fact use two different variants as benchmarks.

In the high parton energy limit, upon assuming weak coupling between the energetic
parton and the medium
the dominant mechanism of energy loss is the radiation of nearly collinear gluons 
from the energetic parton that is induced by interactions between the parton and the medium.
If the medium is sufficiently large that many gluons are radiated from the propagating 
parton, the energy loss rate for a parton in representation $R$ is given, to
leading logarithmic accuracy, by \cite{Baier:1998kq}
\be
\label{eq:Rad}
\frac{dE}{dx}= -\alpha_s \frac{C_R}{2} \,\hat q\, x \ ,
\ee
with $\alpha_s$ and $C_R$ being the strong coupling
constant and the Casimir of the parton, and where the jet quenching parameter
$\hat q$  is the transverse momentum squared picked up by the parton per
distance travelled.
While the expression (\ref{eq:Rad}) describes energy loss in the limit in which 
many gluons are radiated,
in most phenomenological applications of radiative energy loss it is assumed 
that a finite number of hard gluons are emitted from the the energetic partons 
and Eq.~(\ref{eq:Rad}) describes the average over many partons with a fixed energy.

By dimensional analysis, the jet quenching parameter $\hat q\propto T^3$.  
For a very weakly coupled plasma at exceedingly high temperatures,
temperatures such that leading order, leading logarithm, 
perturbative computations are trustworthy, the
jet quenching parameter is given by~\cite{Burke:2013yra}
\be
\label{eq:qhatpert}
\hat q = C_A \alpha_s m^2_D T \log B_{\rm rad}\, ,
\ee
where $m_D^2=g^2 T^2 (2 N_c+N_f)/6$ is the square of the Debye screening length of weakly
coupled quark-gluon plasma 
with $N_c$ colors and $N_f$ flavors,
and $B_{\rm rad}$ is a 
jet-energy-dependent regulator that cuts off large momentum transfers to the plasma.  
A regulator is necessary because in a weakly coupled plasma $\hat q$ diverges
logarithmically with the jet energy $E$.
The precise value of $B_{\rm rad}$ 
is not currently known, although some authors 
estimate it to be $B_{\rm rad}\approx1+6E T /m^2_D $. (See Ref.~\cite{Burke:2013yra} for an extensive 
discussion of estimates of the value of $B_{\rm rad}$ and hence
$\hat q$ in different approximations.)
  We shall ignore all logarithms, lumping
them into a prefactor that we shall denote by $\aR$, with the subscript referring
to `Radiative', and write
\be
\frac{dE}{dx}= - \aR \frac{C_R}{C_F}\, T^3\, x\ ,
\label{eq:Rad2}
\ee
with $C_R/C_F=1$ for an energetic quark and $C_R/C_F=9/4$ for an energetic gluon. 
Although below we shall treat $\aR$ as a parameter to be fit to data, before we go
on we should estimate its value in a weakly coupled plasma using the
leading logarithmic order perturbative calculation, 
which we denote
by $\aR^{\rm pert}$.
Combining Eqs.~(\ref{eq:Rad}) and~(\ref{eq:qhatpert}), 
we obtain
\be
\label{eq:aRpert}
\aR^{\rm pert}=2 \pi C_F C_A \left(\frac{2 N_c + N_f}{6}\right) \alpha_s^3 \log B_{\rm rad}\, .
\ee
For later reference, we may evaluate this expression for typical values of the strong coupling 
constant $\alpha_s=0.2-0.3$, as utilized in fits to the data  in Ref.~\cite{Burke:2013yra}, obtaining
\be
\aR^{\rm pert} \sim (0.3-1.0) \log B_{\rm rad} \sim (2-6) \,,
\label{eq:aRpert2}
\ee
where in the second equality we have used the expression for $B_{\rm rad}$ given
above for jets with energy $E=100$~GeV in a plasma with temperature $T=300$~MeV. 
Note that this logarithm is large, which suggests that, even 
for the high energy jets at the LHC,  leading logarithmic expressions 
such as Eqs. (\ref{eq:Rad})  and (\ref{eq:qhatpert}) are inapplicable and a resummation, as advocated 
in Refs.~\cite{CasalderreySolana:2007sw,Iancu:2014kga,Blaizot:2014bha}, may be needed.

The expression (\ref{eq:Rad2}) captures the leading $x$ and $T$ dependence of radiative 
energy loss at weak coupling. 
We shall treat $\aR$ as a free parameter, fitting it to one
piece of experimental data and then asking how a model in which we use
the expression (\ref{eq:Rad2}) to describe the energy loss of the partons in
a shower fares in comparison to other data.

One reason why it makes sense to treat $\aR$ as a parameter to be fit to data
is that not all of the energy radiated from the initial parton corresponds to 
jet energy loss.  At emission, the radiated gluons are nearly collinear with
the energetic parton, meaning that if the gluons are energetic enough
they remain part of the jet.  This corresponds to 
medium modification
of the branching probability within the shower, without significant energy loss from the jet cone.
However, the subsequent rescattering and further splitting of the radiated gluons
can serve to rapidly soften the gluons, and decorrelate their directions with that
of the energetic parton.  This decorrelation between the directions of the radiated gluons
and the jet direction is expected to be most efficient for the softer radiated gluons and
less efficient for the harder radiated gluons~\cite{CasalderreySolana:2010eh}.
What this means is that the $\aR$ that
we need should be smaller than that obtained in the perturbative calculations,
smaller by a factor that is at present hard to estimate.

Note that we do not propose our simplified approach as a competitor to
more sophisticated Monte Carlo methods  for analyzing the effects of radiative
energy loss on jets being developed by others~\cite{Lokhtin:2008xi,Renk:2008pp,Schenke:2009gb,Armesto:2009fj,Lokhtin:2011qq,Zapp:2013vla}.  
It is in fact clearly inferior, since
we do not track the radiated gluons, treating them as `lost'.  This approach makes
sense in our hybrid model, where the lost energy rapidly becomes soft thermal radiation.
It does not make sense quantitatively here.  Our goal  is solely to have
a benchmark against which to compare our hybrid model.

Finally, and with the aim of exploring the sensitivity of different observables to the path-length dependence of the jet energy loss, we will study a somewhat more extreme model for  energy loss at weak coupling in which 
we assume that $dE/dx$ is given by a collisional rate. 
Collisional energy loss is subdominant to radiative energy loss at weak coupling in the high parton
energy limit, and for this reason it is neglected in many studies.
However, it has been pointed out~\cite{Wicks:2005gt} 
that, while subdominant, these processes play an important role, especially for heavy quarks moving through the plasma.  
Here, we shall not advocate any underlying dynamical picture on the basis
of which to justify including collisional processes.  What we shall do, simply, is to introduce
a third model in which, like in our hybrid model,
parton branching within the shower proceeds as in vacuum and in 
which the energy loss of each parton in the shower is given by
the collisional energy loss rate in a weakly coupled plasma, whose
parametric dependence takes the form~\cite{Wicks:2005gt}
\be
\frac{d E}{dx} = - \aC \frac{C_R}{C_F} T^2\ ,
\label{eq:Coll}
\ee
where we treat $\aC$ (this time the subscript signifies ``Collisional'') as
a fit parameter
 to be constrained by one piece of experimental 
data.
This expression captures the leading temperature, energy and 
path length dependence of the perturbative collisional rate.
For an ultra-relativistic parton
in a weakly coupled plasma, $\aC$
is given to leading logarithmic order in perturbation theory
by~\cite{Wicks:2005gt}
\be
\label{eq:aCpert}
\aC^{\rm pert}= C_F \pi \alpha_s^2 \left(\frac{2 N_c + N_f}{6}\right) \log{B_{\rm coll}} \,.
\ee
where, as before, $B_{\rm coll}$ regulates the effect of large momentum transfer scatterings in the 
medium and is understood to be proportional to the parton energy. The precise expression for $B_{\rm coll}$ 
depends on the criteria used in the 
regularization; see Ref.~\cite{Wicks:2005gt} for a compilation of expressions from the literature. 
As in the case of radiative energy loss, we can substitute
$\alpha_s=0.2-0.3$ into (\ref{eq:aCpert}) and estimate the
value of $\aC$ if we assume that these values of $\alpha_s$ are small
enough for a leading logarithmic calculation to be relevant, obtaining
 \be
 \aC^{\rm pert} \sim (0.25-0.6) \log{B_{\rm coll}} \sim 1.6 -3.3 \,, 
 \label{eq:aCpert2}
 \ee
where in the second equality, we have used $B_{\rm coll}= 6 E T/m^2_D$. 
As in the case of radiative energy loss, the logarithmic factor is large
which means that it is doubtful that these values of $\alpha_s$ 
are small enough for these leading logarithmic expressions
to be reliable.

We have chosen the ratio of Casimirs appearing in both Eqs.~(\ref{eq:Rad2}) 
and (\ref{eq:Coll}) such that 
the parameter $\aR$ (or $\aC$) that
we shall obtain by fitting our expressions for the radiative (or collisional)
energy loss to data is that for the energy loss of a quark
moving through a weakly coupled plasma,
while a gluon gets an additional factor.   Note that the dependence
of Eqs.~(\ref{eq:Rad2}) and (\ref{eq:Coll}) 
on $C_A/C_F$ is much stronger than that in
(\ref{eq:Elrate}), obtained at strong coupling.   
We will return
to this important distinction between energy loss in a strongly
coupled plasma and that in a weakly coupled plasma 
in Section~\ref{cfcadep}.

\section{\label{MC}Monte Carlo Implementation}

The implementation of the hybrid model that we have described requires several steps,  beginning with 
the generation of jets and the modification of their evolution due to energy loss, but also including
the hydrodynamic calculation of the space and time dependence of the bulk medium
created in the heavy ion collision. 
The procedures used for the calculations reported in this work are presented in this Section.

We generate hard processes using PYTHIA 8.170~\cite{Sjostrand:2007gs}.\footnote{
After most of work presented in this paper was completed we became aware that this version of PYTHIA suffers 
from a bug which affects the description of hadronization.  Since we will work at the partonic
level throughout, this bug has no effects on our results. We have explicitly checked
this by recomputing some of our results using PYTHIA 8.183.}
Since at the LHC center of mass energy and in the range of momentum relevant 
for our analysis ($\pt \sim \mathcal{O} (100 {\rm GeV})$), the modification of the nuclear parton distribution functions with respect to the proton ones is very small~\cite{Eskola:2009uj,Chatrchyan:2014hqa}, we simulated 
high energy jet production in proton-proton collisions at $\sqrt{s}= 2.76$ TeV. 
Since these events are later embedded into a hydrodynamic model for the bulk matter produced in the 
nucleus-nucleus collisions, we do not include
the underlying event in the PYTHIA treatment of the proton-proton collision in our calculation.
We use the PYTHIA $\pt$-ordered shower  to evolve the hard process 
from the initial virtuality down to a typical hadronic scale of $Q_0=1~{\rm GeV}$, at which we stop the evolution. 
At this scale, vacuum event generators 
switch to phenomenological models of hadronization, like 
the Lund string model which is incorporated into PYTHIA. 
For a number of reasons, the
nonperturbative hadronization process is expected to be altered in a heavy ion collision
relative to that in vacuum.  For example, most of the soft hadrons in a heavy
ion collision will be formed via the
coalescence of quarks and gluons from the expanding and cooling plasma rather
than directly from partons produced initially and their fragments~\cite{Fries:2003vb,Molnar:2003ff}.
Furthermore, even if we
only look at hadrons that are formed via fragmentation, hadronization in
this setting is still modified by the presence of the medium via changes in how
color flows~\cite{MehtarTani:2010ma,MehtarTani:2011tz,CasalderreySolana:2011rz,Beraudo:2011bh,Beraudo:2012bq,Aurenche:2011rd}. 
In order to avoid complicating the interpretation of our results with currently
unconstrained hadronization dynamics, throughout this paper we will work  at the
partonic level and focus on observables that are less sensitive to the hadronization process. 
For example, in jet observables these corrections are, at least in vacuum, smaller 
than $10\%$ \cite{Soyez:2011np}.

On an event by event basis, the events generated by PYTHIA each
initiate a decay chain which will be the starting point for our implementation of medium effects. 
As we have argued in Section \ref{aha}, 
in our hybrid model we shall neglect the possibility that the presence of the medium may
result in modification of the splitting probabilities, or modification to the
locations in space and time where splitting occurs obtained via Eq.~(\ref{eq:time}).
We are neglecting the fact that the reduction in the available energy due
to the loss of energy of a parton in the shower
leads to a reduction in the phase space available when that parton subsequently
splits.
In this exploratory study we will neglect such phase space effects and assume that the overall structure of the decay chain remains the same even after we make the partons in the shower lose energy.

We place the point of origin of each of the dijet processes  generated by PYTHIA
in the plane transverse to the collision axis 
at a location selected with a probability proportional to the number
of binary collisions at that location in the transverse plane.  The
showers generated by the dijets proceed in space and time according
to Eq.~(\ref{eq:time}), propagating outward along their (randomly
selected) direction of motion.
 Since the dijet production process is hard, dijets are produced very early ($\tau\sim 1/Q$),   
 prior to the proper time at which the plasma produced in the collision hydrodynamizes,
 $\tau_{\rm hydro}$. 
 We will assume that during 
 the short proper time before $\tau_{\rm hydro}$, 
 the jets propagate unperturbed.\footnote{This is an assumption that could 
 be improved upon in future, once the analysis of the early pre-equilibrium
 energy loss of heavy quarks in Ref.~\cite{Chesler:2013cqa} is extended to light quarks.
 That analysis indicates that energy loss sets in only after a delay time of order $1/(\pi T)$
 after the moment during the collision when the energy density is at its maximum,
 $T$ being the temperature at the time of hydrodynamization.  
 In addition, the analysis of the collision of sheets of energy density in Ref.~\cite{Casalderrey-Solana:2013aba}
 indicates that if the sheets are thin enough there is 
 a prior delay of order $1/(\pi T)$ between the collision time and the 
 time when the energy density peaks.
  The 
 results of Refs.~\cite{Chesler:2013cqa,Casalderrey-Solana:2013aba} together indicate
 that there will surely be some energy loss before $\tau_{\rm hydro}$ but that it is not
 expected to be large.}
   After $\tau_{\rm hydro}$, the jets encounter the hydrodynamically 
 expanding plasma and the different fragments of the jet suffer energy loss,
 according to (\ref{eq:Elrate}) in our hybrid strong/weak coupling model
 or according to (\ref{eq:Rad2}) or (\ref{eq:Coll}) in our models of 
 weakly coupled radiative or collisional energy loss.
To determine the local properties of the plasma at the position of the fragments, we 
embed the jet shower into 
the boost-invariant ideal hydrodynamic simulations
of the expanding cooling plasma produced in heavy ion
collisions with $\sqrt{s}=2.76$~TeV per nucleon that we have obtained from
Ref.~\cite{hiranoLHC}. 
 These simulations reproduce 
 the multiplicity of charged particles produced at mid-rapidity
 at the LHC.\footnote{It would be interesting
to repeat our analysis using a three-dimensional viscous hydrodynamics
simulation, ideally one that includes event-by-event fluctuations in the initial state at the
time of hydrodynamization.  We leave this to future work.}
Since in the  simulations of Ref.~\cite{hiranoLHC} 
the hydrodynamic fields are initialized at $\tau_{\rm hydro}=0.6$~fm, we will take this 
as our hydrodynamization time.   
From these simulations we determine 
the temperature of the plasma at each point in space and time, and hence the 
spacetime-dependent temperature
that each parton in the fragmenting shower encounters on its way through and eventually out
of the expanding, cooling, droplet of plasma. We use this
spacetime-dependent temperature to integrate the different expressions for the energy loss rate $dE/dx$
discussed in Section \ref{elm} over the path of each parton in the shower during its lifetime, {\it i.e.}~from
the time when it is produced in a splitting process to the time when it itself splits.

The procedure described above assigns an energy loss to each of the virtual partons in the shower. 
However, it does not determine how the lost energy is distributed among the 
several particles that are produced when each virtual parton splits, or decays.
Consistent with the assumption that  the medium does not change the splitting probabilities in the shower, 
since these splitting probabilities depend on the energies of the daughter partons 
only through the fraction of the parent parton energy that each daughter obtains
as a result of the splitting we choose to distribute the energy lost by the parent
parton as a reduction in the initial energy of each of the daughters according to this fraction.
As  they themselves propagate through the medium subsequently, these decay partons loose additional 
energy until they split again. 
Therefore, 
the total energy lost by a particular final parton that escapes from the medium depends on the 
detailed history of splitting and propagation that led to that parton. 

Since the goal of this work is to study the effect on high energy jets of energy loss in strongly coupled plasmas, we will not describe the degradation of the jet energy in the hadron gas produced after 
the plasma cools through the QCD phase transition at $T\sim T_c$.  We focus only on
the energy loss as the jet propagates through the strongly coupled plasma
with $T>T_c$.
To ensure that we do not apply the strong coupling results to the late time resonance gas, we will stop the computation of energy loss when the temperature of the system falls below $T_c$, which we identify 
with the crossover
temperature of the QCD transition that separates the plasma from the hadron gas.
Since the 
QCD transition is a cross-over, 
$T_c$ is not sharply defined and
its precise  value depends on the procedure used to determine it. 
The hydrodynamic simulations from Ref.~\cite{hiranoLHC} that we are using
employ an equation of state obtained from the lattice QCD calculations in Ref.~\cite{Bazavov:2009zn}.
Although more recent lattice calculations favor a slightly lower value of $T_c$, since
we are obtaining the temperature profile from hydrodynamic calculations done
according to the QCD thermodynamics of Ref.~\cite{Bazavov:2009zn} we
will vary $T_c$ in the range $180< T_c < 200$ MeV specified in Ref.~\cite{Bazavov:2009zn}.
We shall employ this variation in our choice of $T_c$ as a device with which to
estimate the systematic uncertainty in the results that we obtain from the computations that we shall perform using our
hybrid model.

Finally, in addressing RHIC data we will employ an identical procedure except that we 
start with hard dijets produced (by PYTHIA) in collisions with
$\sqrt{s}=200$ GeV per nucleon and we replace the hydrodynamic profile for LHC collisions
with that for RHIC collisions, also
obtained from
Ref.~\cite{hiranoLHC}.

In the next Section, we describe how we reconstruct the jets in our hybrid model and
compare them, in various ways using various measured observables, to jets reconstructed
from heavy ion collision data.

\section{\label{cd}Comparison with Jet Data}

We have described the implementation of our hybrid model in full detail in
the two previous Sections.  All that remains is to choose the one dimensionless free parameter
$\aSC$, defined in Eq.~\ref{eq:xstop}, 
that we have introduced into our description of the energy loss of an individual
parton in the PYTHIA shower as it propagates through the strongly coupled plasma
and the model will then be fully specified.
As explained in Section \ref{elm}, we are assuming that  the strongly coupled dynamics fixes 
the parametric dependence of the energy loss rate $dE/dx$, given in Eq.~(\ref{eq:Elrate}),
and the stopping distance $x_{\rm stop}$, given in Eq.~(\ref{eq:xstop}),
but not the overall normalization
of $x_{\rm stop}$. 
Therefore, our model possesses one free parameter, which we need to fit to data. 
Once this has been done, we will be able to 
study different jet observables and extract the effect of the medium on each of them.

\subsection{Jet reconstruction and jet $\Raa$}

The first observable that we shall compute is $\Raa$ for jets, as a function of $p_T$, the 
transverse momentum of the jet, and as a function of the centrality of the heavy ion
collision.\footnote{The ``centrality'' of a collision between heavy ions refers to its 
impact parameter.  Nearly head-on collisions, with the smallest impact parameters,
are referred to as central collisions; peripheral collisions, with large
impact parameter, are noncentral.   The impact parameter is not directly measured,
but it is nevertheless possible to bin heavy ion collision data as a function of
impact parameter, for example using the fact that the total number of hadrons   
produced in a heavy ion collision is anticorrelated with the impact parameter
of the collision.  Central collisions have the highest multiplicity; peripheral collisions
the lowest.  Experimentalists therefore bin their events by multiplicity, using
that as a proxy for the impact parameter.  The terminology used refers, for example,
to the ``0-10\% centrality bin'' and the ``10-20\% centrality bin'', meaning the 10\%
of events with the highest  multiplicities (and lowest
impact parameters) and the next 10\% of events with the
next highest multiplicities (and next lowest impact parameters).  The correlation between event multiplicity
and impact parameter is described well by the Glauber model of multiple
scattering~\cite{Bialas:1976ed,Miller:2007ri}, which relates the event multiplicity to the number
of nucleons that participate in the collision ($N_{\rm part}$) which in turn
can be related via a geometrical calculation to the impact parameter of the
collision.  In our calculations, we take the tabulation of the range of 
impact parameters that corresponds to a given centrality bin defined
via the multiplicity distribution for Pb-Pb collisions at the LHC from
 Ref.~\cite{hiranoLHC}. When we distribute the points of origin
 of our PYTHIA jets in the transverse plane, we do so 
with a probability distribution for
 the impact parameter $b$ 
 dictated by the number of collisions at each $b$
 within the range corresponding
 to a given centrality bin. 
 In order to then
 apply our energy loss prescription to the partons in the PYTHIA
 shower, we embed the PYTHIA jet in the hydrodynamic solution
 from Ref.~\cite{hiranoLHC} corresponding to the mean value of
 the impact parameter in the interval associated with the given
 centrality bin.
}  
The jet $\Raa$ is the ratio of the number of reconstructed jets with a given
$p_T$ that we find in heavy ion collisions in a given centrality bin to the number of
jets with that same $p_T$ in $N_{\rm binary}$ proton-proton collisions with the
same $p_T$, where $N_{\rm binary}$ is the number of proton-proton collisions that
occur in a heavy ion collision of the given centrality, according to a Glauber 
model.
Because the production cross-section for jets is a rapidly falling function of $p_T$, if the jets
in a heavy ion collision have lost energy due to the passage of the partons in the jet
through the strongly coupled medium this results in $\Raa<1$.
To determine the prediction of our model for the jet $\Raa$, we need to
reconstruct jets both in heavy ion collisions within our model (as described in previous sections,
including the effects of energy loss) 
and in proton-proton collisions as described
by PYTHIA with the underlying event switched off, as explained in Section~\ref{MC}.
To obtain the principal results of this paper, we generated 300,000 PYTHIA events 
with $p_T$ greater than a cut that we set to
50 GeV for collisions with centralities in each of four ranges (0-10\%, 10-30\%, 30-50\% and 50-70\%).   
We varied the $p_T$ cut to make sure 
that the jet spectrum in the (higher) range of $p_T$ where we 
performed our analyses was insensitive to the value of the cut.
We used the PYTHIA events without modification
to describe jets in proton-proton collisions.
As described in Sections~\ref{elm} and \ref{MC}, to describe
quenched jets in heavy ion collisions we embed the PYTHIA events
in a hydrodynamic description of the matter produced in a heavy
ion collision and apply our prescription for energy loss to each
parton in the PYTHIA shower.
We then analyze 
the output of our model calculations of quenched jets in heavy ion collisions
and of proton-proton jets using FastJet~\cite{Cacciari:2011ma}, 
with which 
we reconstruct jets using the
anti-$k_t$ algorithm~\cite{Cacciari:2008gp}. 
Defining a jet, via any reconstruction algorithm, requires the specification
of a resolution parameter, $R$.  
This parameter can be understood as the opening angle (in radians) of the jets we reconstruct,
 although 
the precise meaning of $R$ is different for different reconstruction algorithms. 
We shall set the reconstruction parameter in the anti-$k_t$ algorithm to
$R=0.3$ for Pb-Pb collisions at LHC energies and to $R=0.2$ for Au-Au
collisions at RHIC energies because we shall compare the predictions of
our model to jet measurements from LHC and RHIC data that 
employ these values of $R$.
As we have discussed in Section~\ref{MC}, the output of our model 
is partons not hadrons, and we are reconstructing jets from those partons.
For this reason, we will focus on jet observables that
are relatively
insensitive to details of the hadronization process.

\begin{figure}[tbp]
\centering 
\begin{tabular}{cc}
\includegraphics[width=.5\textwidth]{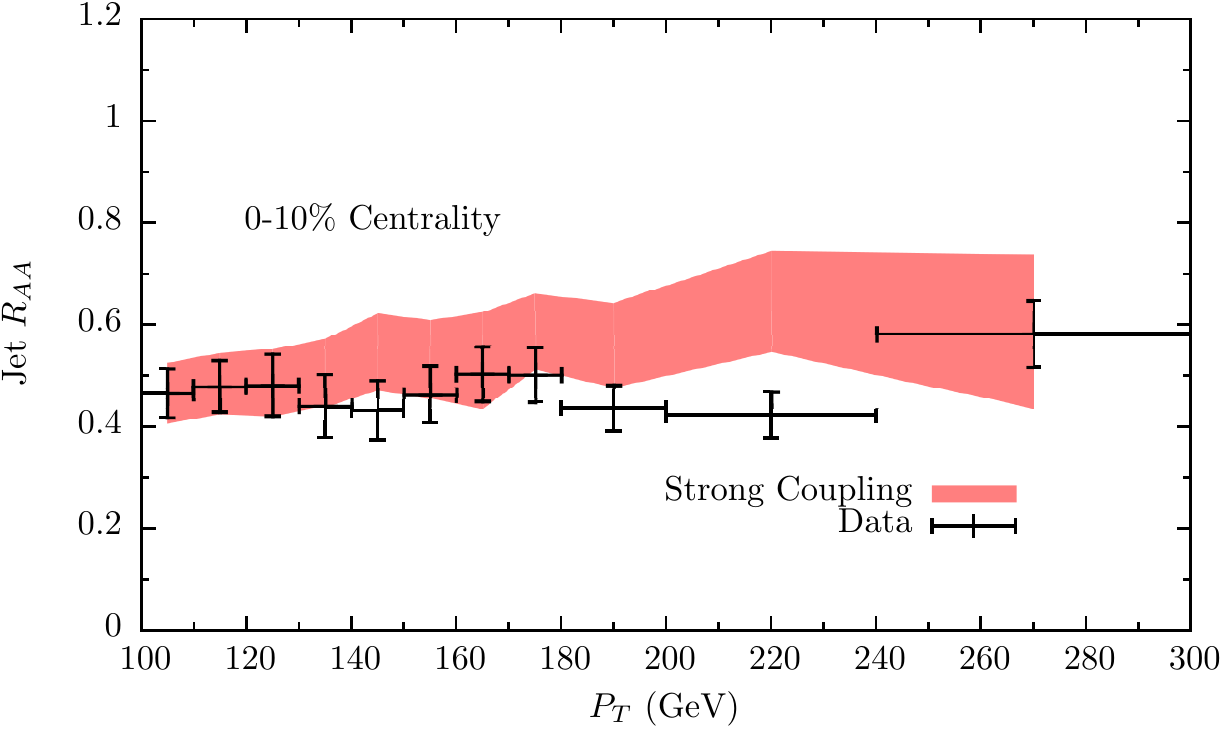}
&
\includegraphics[width=.5\textwidth]{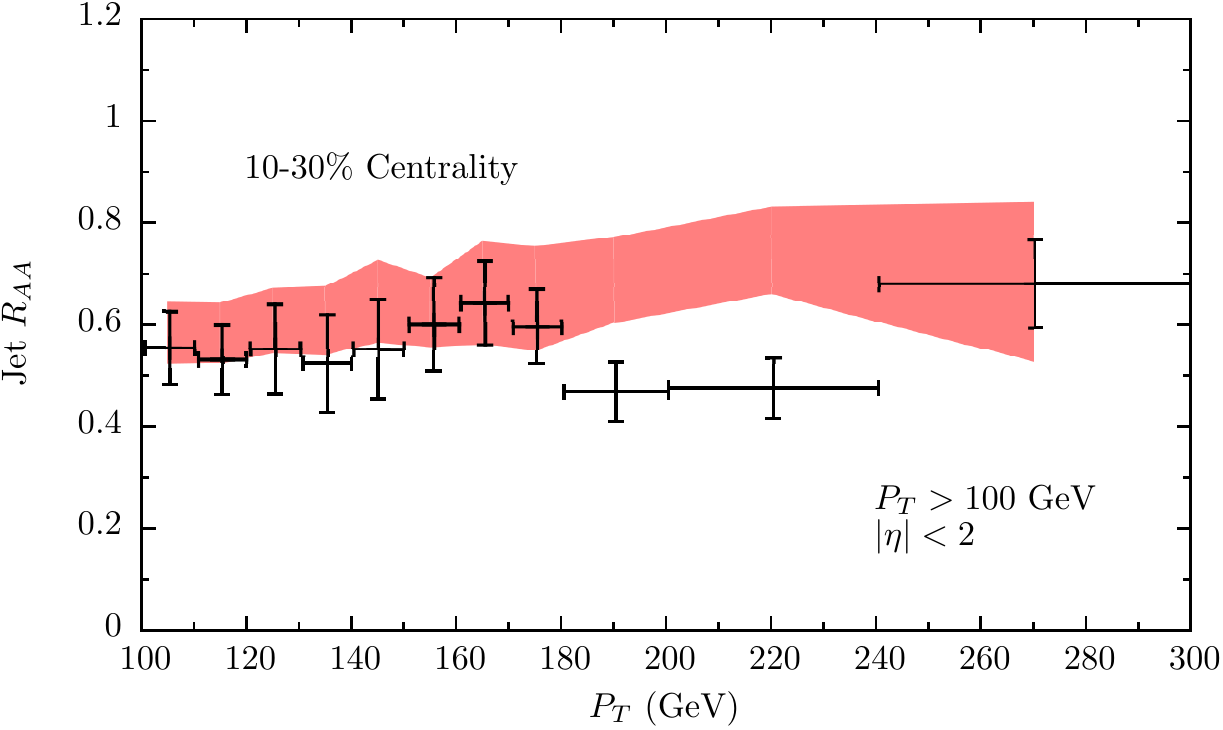}
\\
\includegraphics[width=.5\textwidth]{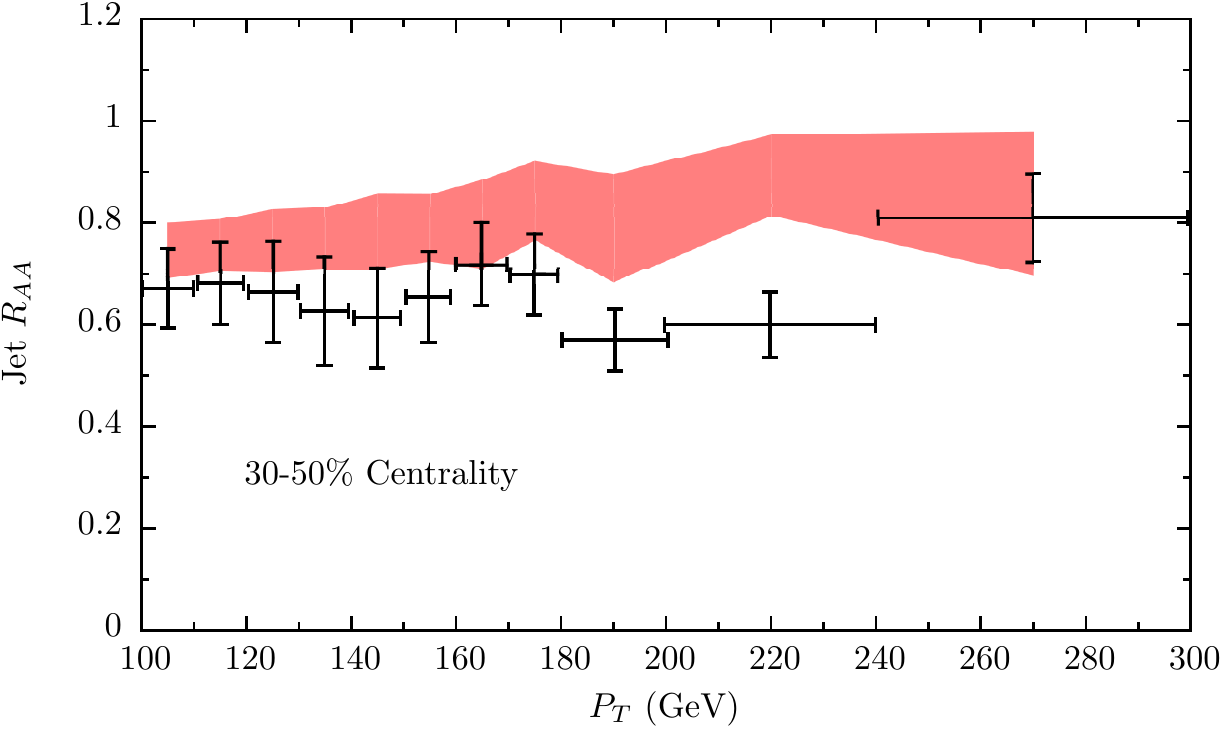}
&
\includegraphics[width=.5\textwidth]{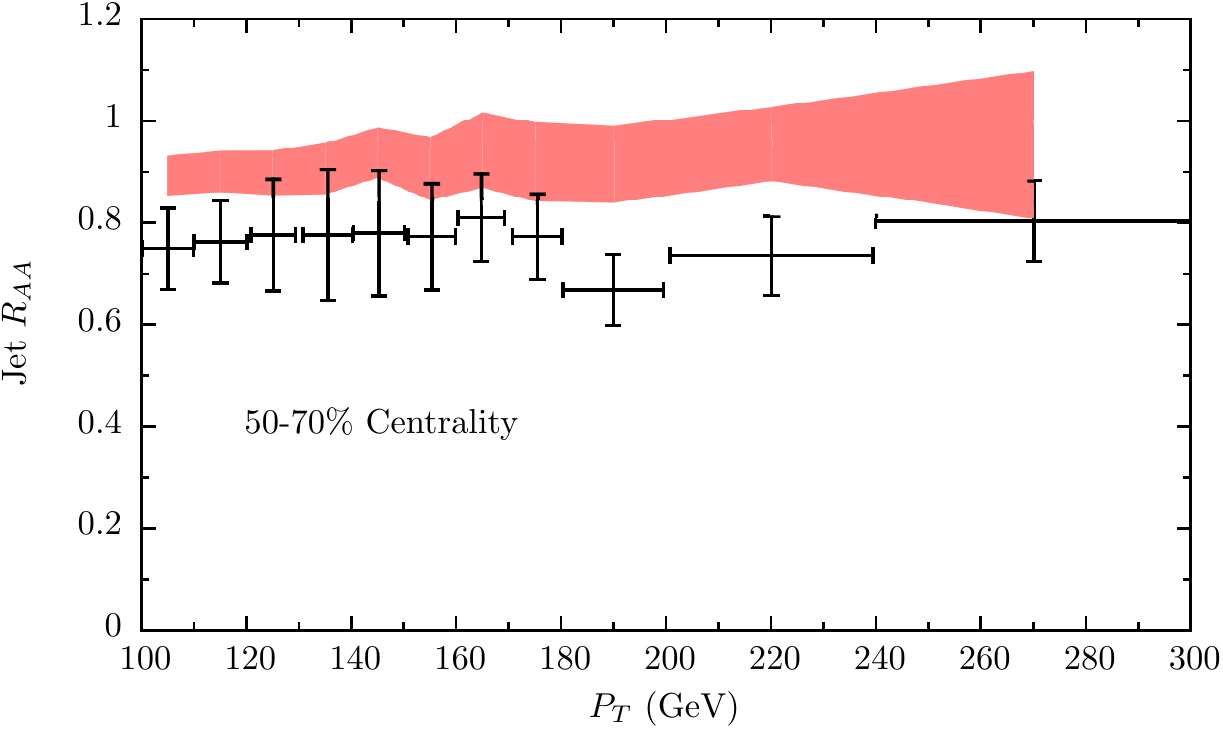}
\end{tabular}
\caption{\label{Fig:RAA}  Jet $\Raa$ as a function of $\pt$ for different centralities in our
hybrid model for jet quenching in strongly coupled plasma  
compared to preliminary CMS data from Ref.~\cite{Raajet:HIN}. 
The results of our calculations in our hybrid strongly coupled model,
shown by the colored bands, are completely specified
once we have fixed the one free parameter in the model so that the model
agrees with the left-most data point in the top-left panel, namely the jets with $100~{\rm GeV}< \pt<110~{\rm GeV}$
in the most central collisions.  Once this point has been fitted, the $\pt$ dependence
and centrality dependence of $\Raa$ are outputs of the model.
}
\end{figure}

In order to fit the value of the one free parameter $\aSC$ in our parametrization (\ref{eq:Elrate}) and (\ref{eq:xstop}) 
for the rate of energy loss $dE/dx$ of each of the partons in the PYTHIA shower, we
calculate $\Raa$ for jets with $100 \leq p_T\leq 110$~GeV
in the range of pseudorapidities $-2 \leq \eta \leq 2$ in 
 the 0-10\% most central Pb-Pb collisions at LHC energies, with 
 collision energy $\sqrt{s}=2.76$~TeV per nucleon pair.  
This quantity has been measured by the CMS collaboration,
and in the data it lies between 0.42 and 0.51.  
(For this and for all experimental data quoted in our paper,
we have added the statistical and systematic errors in quadrature.) 
We find
that we can reproduce this measured result with our model
as long as we choose $\aSC$ between 0.26 and 0.35.
In determining this range of allowed values of the parameter $\aSC$
we have included the theoretical uncertainty in the critical temperature $T_c$,
discussed in Section~\ref{MC},
as well as the uncertainty that enters via the uncertainty in the experimentally
measured quantity.  The latter dominates the uncertainty in the extracted
value of $\aSC$.  Henceforth, in all our plots we will show a band of results
obtained from our model corresponding to varying $\aSC$  between 0.26 and 0.35,
a range that incorporates both experimental and theoretical uncertainty.

With $\aSC$ now fixed, the first results that we obtain from our model are the dependence of the jet $\Raa$ on
$p_T$ and on the centrality of the collision, for Pb-Pb collisions at $\sqrt{s}=2.76$~TeV.
We show our results in Fig.~\ref{Fig:RAA}.    We see that our hybrid model predicts
a jet $\Raa$ that is only weakly $p_T$-dependent, in agreement with the preliminary
CMS data from Ref.~\cite{Raajet:HIN}.  The evolution of the jet $\Raa$ with increasing centrality
is consistent with the data until we get to the most peripheral bin, for which our model predicts
less quenching than is seen in the data.  
 This discrepancy  may be due in part to the fact that we are not including the energy loss in the hadronic phase in our computation, since peripheral collisions will spend less time in the plasma  phase
 making the time spent in the hadronic phase proportionally more relevant.

\begin{figure}[tbp]
\centering 
\includegraphics[width=.49\textwidth]{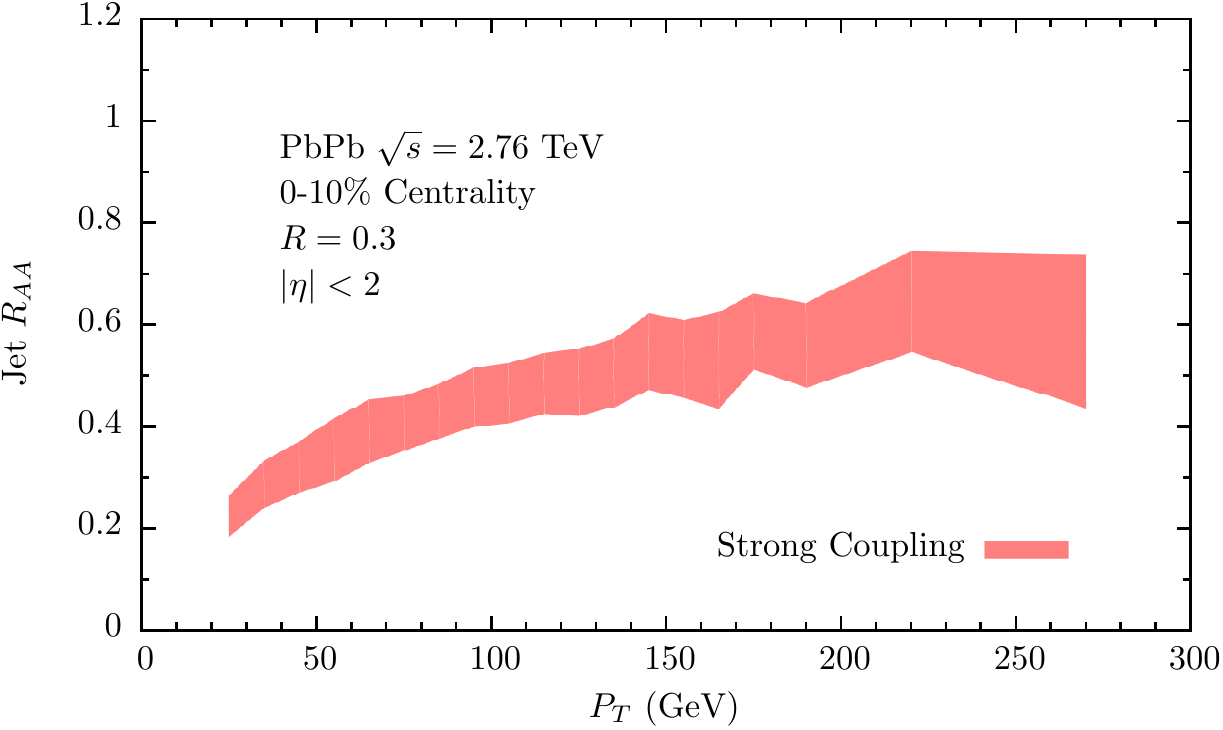}
\includegraphics[width=.49\textwidth]{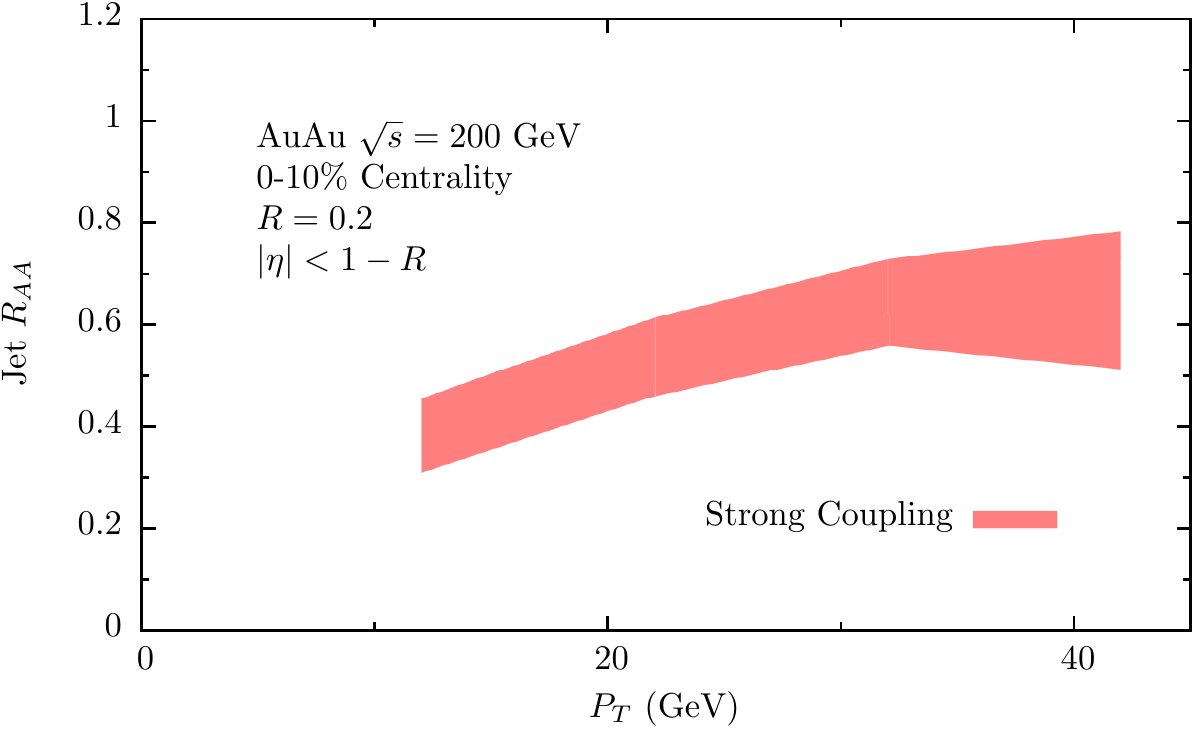}
\caption{\label{Fig:RAApt}   Predictions of our hybrid strongly coupled model for 
jet $\Raa$ as a function of $\pt$ for central Pb-Pb collisions at the LHC
with $\sqrt{s}=2.76$~TeV per nucleon (left) and Au-Au collisions at RHIC with $\sqrt{s}=200$~GeV per 
nucleon (right).  In both cases, we only show our results for collisions in the 0-10\% centrality bin.} 
\end{figure}

In Fig.~\ref{Fig:RAApt} we further explore 
the $\pt$ and $\sqrt{s}$ dependence of the jet $\Raa$ within our hybrid approach.  
In the left panel, we extend our computation of jet suppression
down to 15 GeV for the most central LHC collisions, using a sample
of PYTHIA jets generated with $\pt$ greater than a 10 GeV cut. 
Because the jet production cross-section falls rapidly with $\pt$, in order
to have sufficient statistics over this wide range in $\pt$
we generated several independent samples of jets, 
each with $\pt$ greater than a higher value of the cut than in the
sample before, employing cuts of 10, 35 and 50 GeV.
We then merged each sample with the previous one away from these cuts.
In this way we were able to obtain a sample of jets with reasonable statistics for $\pt$
ranging all the way from 15~GeV to 270~GeV. 
Even over this extended range of $\pt$, the 
jet suppression factor $\Raa$ varies relatively little with transverse momentum.
This is in qualitative agreement with  $\Rcp$ 
measurements by ATLAS \cite{Aad:2012vca} and 
charged jet $\Rcp$ measurements by ALICE \cite{Abelev:2013kqa}, 
which both report suppression measurements down 
to this range of $\pt$ with a similarly weak dependence on $\pt$.
Nevertheless, at present we refrain from a quantitative comparison with these data, for two
reasons.  $\Rcp$ is the ratio of the number of jets with a given $\pt$
in central collisions to an expectation based upon data in peripheral collisions, rather
than an expectation based upon data in proton-proton collisions as in $\Raa$. Given the
disagreement that we see  between our model and the data in
the  peripheral bin at the higher values of $\pt$ 
displayed in Fig.~\ref{Fig:RAA},  we cannot make a direct comparison between
our results at lower values of $\pt$ in Fig.~\ref{Fig:RAApt} and measurements
of $\Rcp$.
And, since we are working at the partonic level, 
we are at present hesitant to compare our results
to measurements of jets defined via charged hadrons only, rather than calorimetrically.
In the right panel of Fig.~\ref{Fig:RAApt}, we repeat  
our analysis for the lower jet energies available in RHIC collisions
with a center of mass energy of $\sqrt{s}=200$~GeV per nucleon,
extending our analysis down to 12 GeV using a sample of PYTHIA
jets generated with $\pt$ greater than a 5 GeV cut.
We chose the jet reconstruction parameter $R=0.2$, as in Ref.~\cite{star:jets}. 
Our results are in good agreement with the preliminary 
experimental measurements reported by the STAR collaboration in Ref.~\cite{star:jets}, at present
still with significant systematic uncertainties.
However, we have again refrained from making a direct comparison since,
as before, it is not easy to compare our partonic jet results with the
charged jet measurements reported in Ref.~\cite{star:jets}. Also, in
making these measurements the STAR collaboration requires the presence
of a semi-hard ($\pt=5-7$~GeV) charged hadron within the jet, a criterion that
is hard for us to reproduce from our partonic computation.

\begin{figure}[tbp]
\centering 
\begin{tabular}{cc}
\includegraphics[width=.5\textwidth]{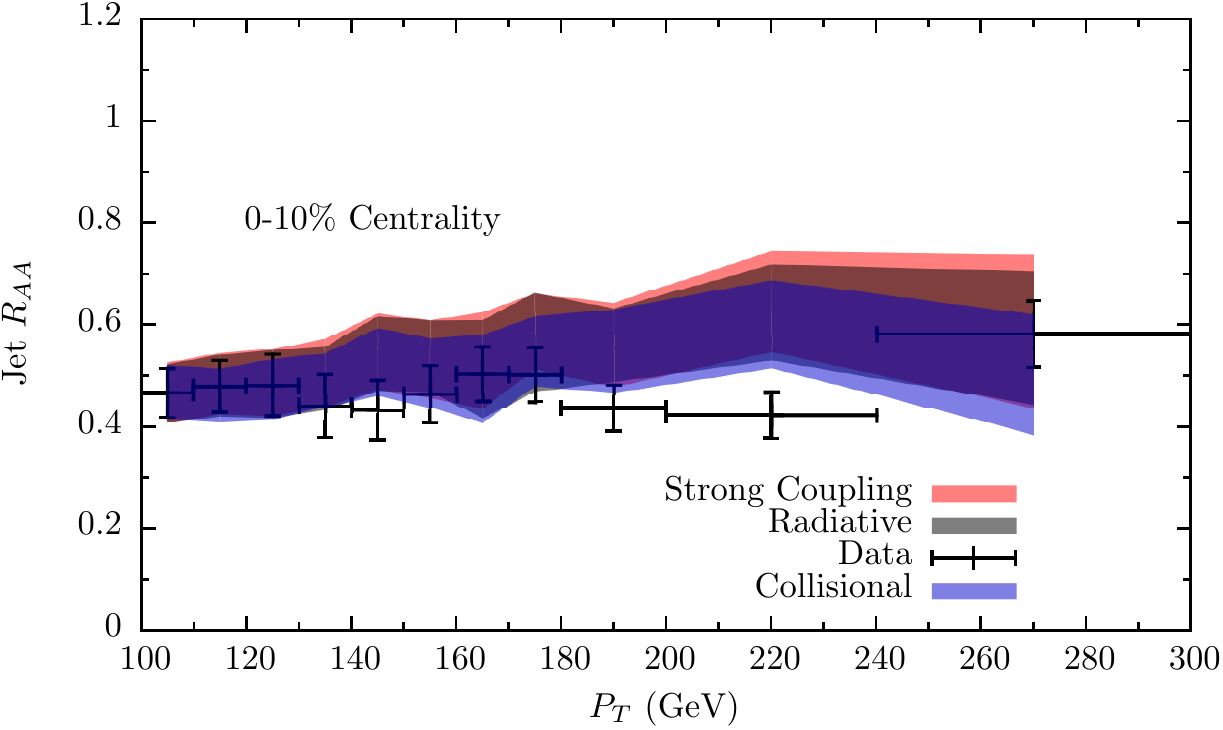}
&
\includegraphics[width=.5\textwidth]{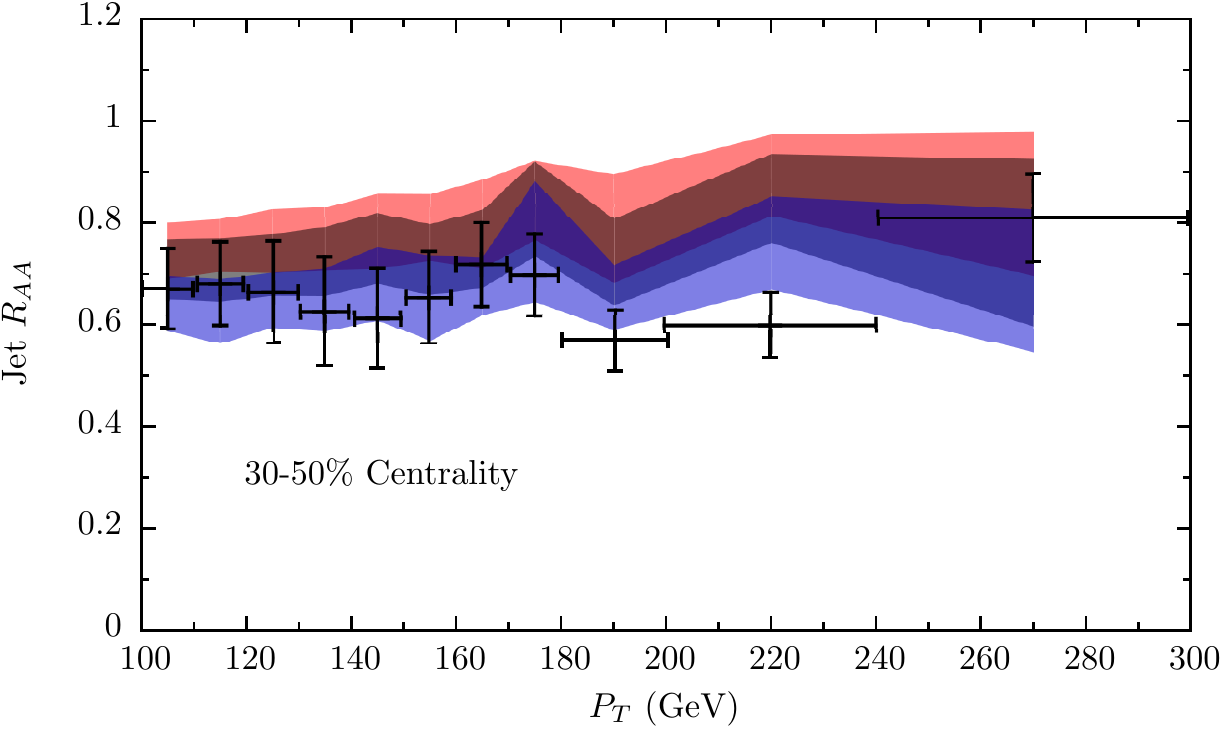}
\end{tabular}
\caption{\label{Fig:RAAcomp}  
Jet $\Raa$ as a function of $\pt$ for LHC collisions
in two different centrality bins for the three energy loss models from
Section~\ref{elm}, as compared to preliminary CMS data \cite{Raajet:HIN}. Each of the three models
for the rate of energy loss $dE/dx$ includes one free parameter, and in each case we have
fitted the value of this parameter to obtain agreement between the model and the data
for $100~{\rm GeV}<\pt<110~{\rm GeV}$ in the most central ($0-10$\%) collisions.}
\end{figure}

The predictions of our model for both
the momentum dependence and the centrality dependence of jet suppression 
are in encouraging agreement with experimental data.
To avoid over-interpreting this agreement, it is important to assess the sensitivity
of the jet $\Raa$ observable to the underlying dynamics of the energy loss.
To gauge this sensitivity, 
we have repeated the analysis for the two other models of the energy loss rate $dE/dx$
described in Section~\ref{elm}. In Fig.~\ref{Fig:RAAcomp}
we show the jet suppression factor $\Raa$ 
in two centrality bins for the strongly coupled (red), 
radiative (grey) and collisional (blue) energy loss models. 
In all three models, as in Fig.~\ref{Fig:RAA} we have fitted the one free
parameter in our description of $dE/dx$ to the left-most data point in the
left panel, finding $0.81< \aR <1.60$ for the parameter $\aR$ defined
in the expression (\ref{eq:Rad2}) for $dE/dx$ in our model for
weakly coupled radiative energy loss
and $2.5 < \aC < 4.2$ for the parameter $\aC$ defined
in the expression (\ref{eq:Coll}) for $dE/dx$ in our model for
weakly coupled collisional energy loss.
Remarkably, despite the fact that the energy dependence and the path-length
dependence of the three different expressions (\ref{eq:Elrate}), (\ref{eq:Rad2}) and (\ref{eq:Coll}) are
very different for the three quite different energy loss mechanisms that we
are modelling, the $\pt$ dependence and the centrality dependence 
of the jet $\Raa$ are quite similar in all three models.

\subsection{Dijet asymmetry}

\begin{figure}[tbp]
\centering 
\begin{tabular}{cc}
\includegraphics[width=.5\textwidth]{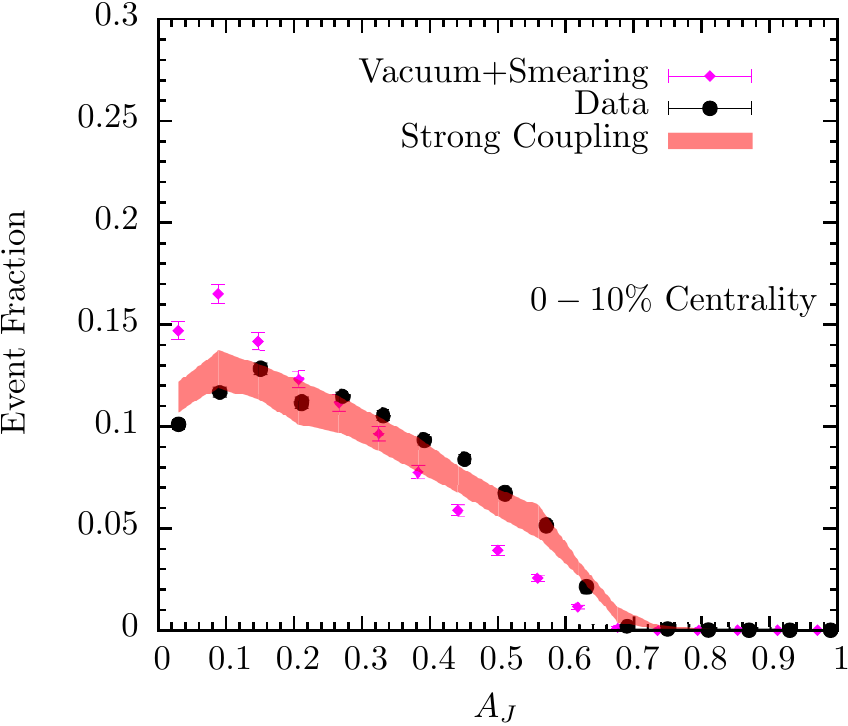}
&
\includegraphics[width=.5\textwidth]{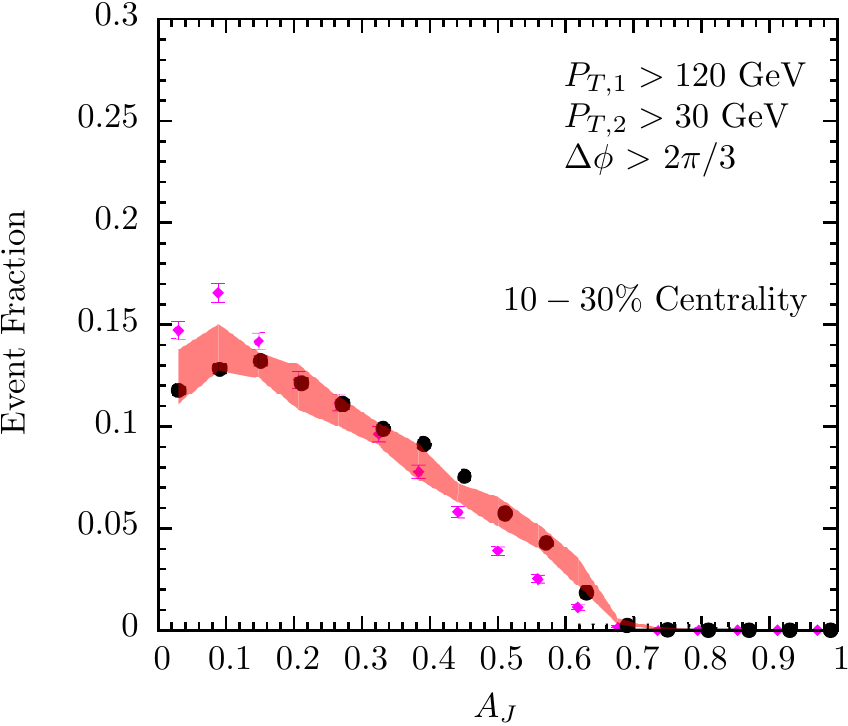}
\\
\includegraphics[width=.5\textwidth]{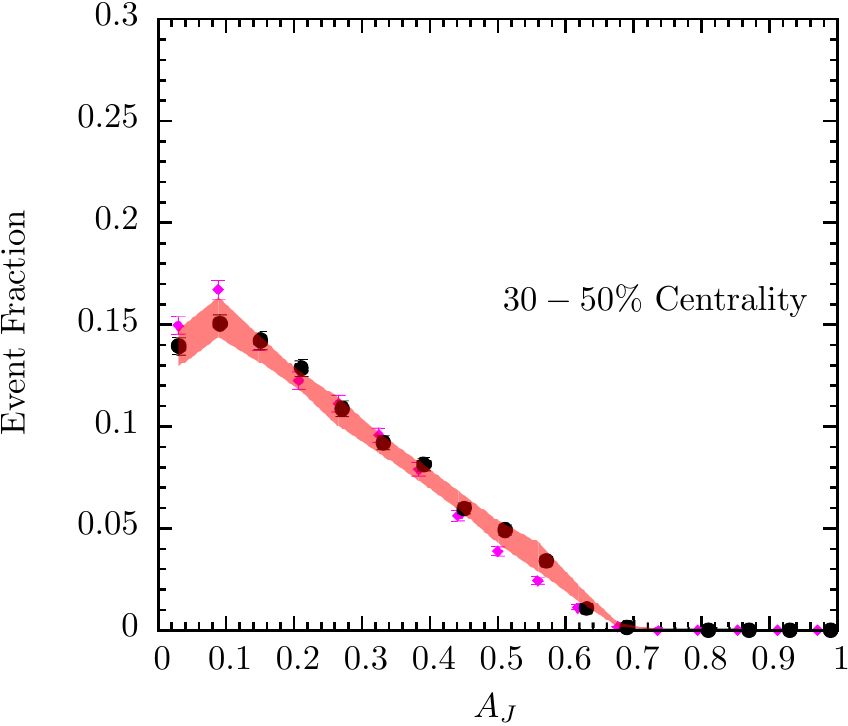}
&
\includegraphics[width=.5\textwidth]{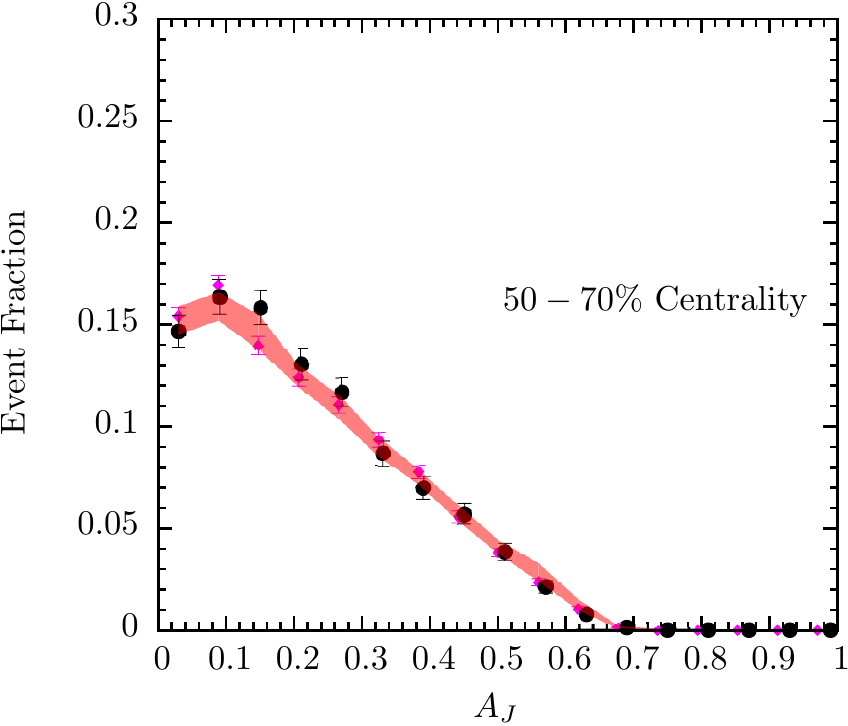}
\end{tabular}
\caption{\label{Fig:Asym}  Red bands show the probability distributions for the dijet imbalance $A_J$
in LHC collisions with four different ranges of centrality predicted by our
strongly coupled hybrid model. The jet momenta are smeared, according to the
prescription given in Ref.~\cite{YYthesis} in order to mimic background subtraction effects. 
Experimental data points are taken from Ref.~\cite{Chatrchyan:2012nia}.
As a comparison, we show the distributions of the dijet imbalance $A_J$ in
the proton-proton collisions that we have obtained from PYTHIA,
including the (centrality dependent) momentum smearing needed in order
to make a fair comparison to the heavy ion results. 
}
\end{figure}

After constraining and then confronting 
the three models with data on the jet suppression $\Raa$, we turn
now to a different jet observable, the dijet imbalance $A_J$~\cite{Aad:2010bu,Chatrchyan:2011sx}. 
Following the data analysis procedure used in the analysis of the experimental data reported in
Ref.~\cite{Chatrchyan:2012nia}, in our Monte Carlo simulation  
we select events containing dijet pairs 
reconstructed with the anti-$k_t$ algorithm  with jet reconstruction parameter $R=0.3$ 
in the pseudorapidity range $\left | \eta \right| <2$
such that the leading jet has $\ponet> 120$~GeV and the subleading jet has $\ptwot> 30$~GeV. 
The asymmetry variable is then defined as $\AJ\equiv (\ponet-\ptwot)/(\ponet+\ptwot)$. 
 Since the data presented by both ATLAS \cite{Aad:2010bu} and 
 CMS \cite{Chatrchyan:2011sx, Chatrchyan:2012nia}  for this observable are not fully 
 unfolded from resolution effects, a direct comparison of the result of our computations with 
 data is not possible.  However, the CMS collaboration has demonstrated that a simple 
 centrality and momentum dependent smearing procedure can reproduce the 
 systematics of such effects, at least for $\gamma$-jet observables, and has
 provided an explicit parameterization for such smearing in that type of measurement~\cite{YYthesis}. 
 Since the corresponding parameterization  for dijet measurements is not yet available, we will use the procedure advocated in  Ref.~\cite{YYthesis} also for dijets.  The result of these computations
is a prediction from our strongly coupled hybrid model for the probability distribution for $A_J$ 
 for heavy ion collisions at the LHC with  four different centrality bins shown in Fig.~\ref{Fig:Asym}. 
 The centrality dependence of the smearing function is illustrated by the violet points which
 show the results of applying the (centrality dependent) smearing to proton-proton events from PYTHIA.
 The energy loss experienced by both jets in the dijet pair tends to increase $A_J$
 in heavy ion collisions, more so in more central collisions.  
 We see this in Fig.~\ref{Fig:Asym}
 as the widening of the asymmetry distribution in more central collisions, both in the predictions
 of our model and in the data. We see from the figure that there is good agreement
 between the predictions of our model and measurements made using LHC data.

\begin{figure}[tbp]
\centering 
\begin{tabular}{cc}
\includegraphics[width=.5\textwidth]{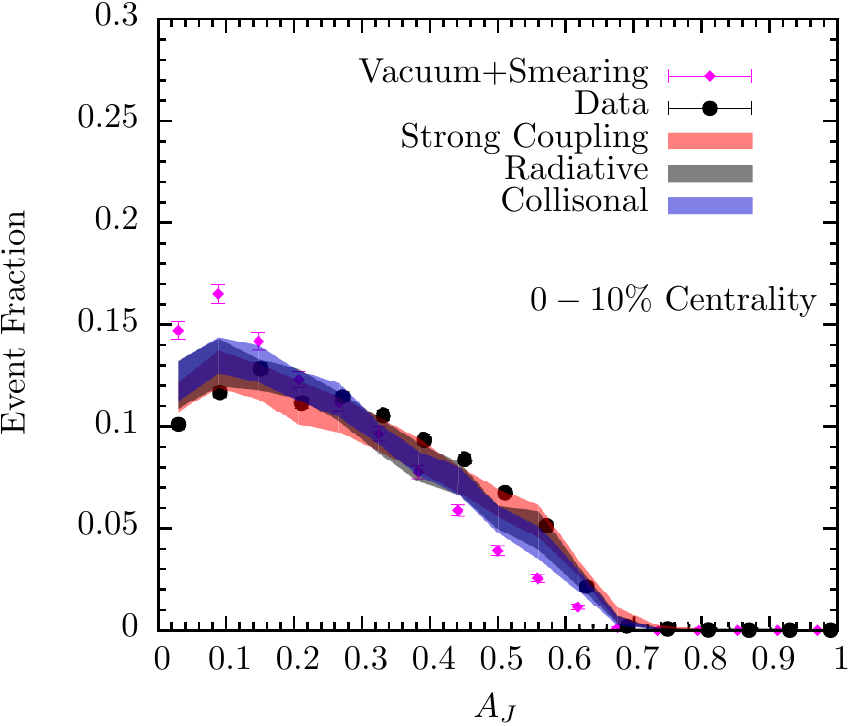}
&
\includegraphics[width=.5\textwidth]{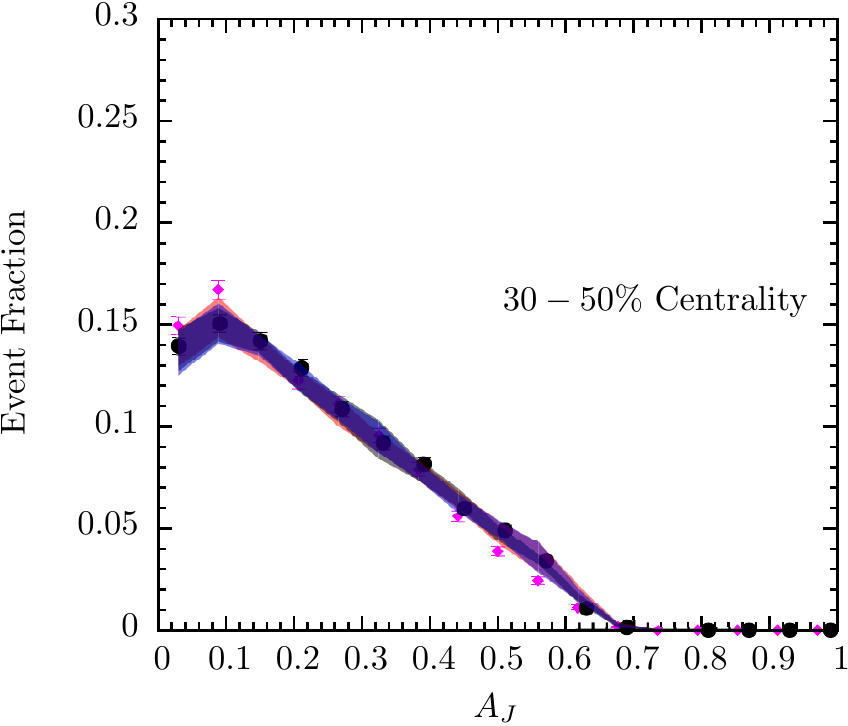}
\end{tabular}
\caption{\label{Fig:Asymcomp}  Probability distribution for the
smeared dijet imbalance $A_J$ for three different models of
the rate of energy loss $dE/dx$ in LHC heavy ion collisions in
two different ranges of centrality.
}
\end{figure}

As before, before over-interpreting the good agreement between the strongly coupled hybrid model
prediction for the dijet asymmetry distribution and the data, in Fig.~\ref{Fig:Asymcomp}
we show the (smeared) results for the dijet asymmetry distribution in events with
two different ranges of centrality if we use the
strongly coupled (red), radiative (grey) and collisional (blue) models for the
rate of energy loss $dE/dx$.
As in the case of the jet suppression $\Raa$, 
our results for the $A_J$ distribution is only weakly dependent on
our choice of the underlying model. Even though the three different models
have quite different path-length dependence for $dE/dx$, 
all three models lead to similar dijet asymmetries.
Although it is a small effect, we do notice here that the strongly coupled model
yields a slightly larger dijet imbalance in the most central collisions 
and that this means it is in somewhat better agreement with the data than the other two models.
Nevertheless, the larger message of Fig.~\ref{Fig:Asymcomp} is the approximate agreement
between the predictions of three models with energy loss rates
that feature very different path-length dependence, indicating that the
these types of jet observables have only limited sensitivity to
the shape of the underlying medium, as observed
previously in Ref.~\cite{Renk:2012cx}.

\begin{figure}[tbp]
\centering 
\begin{tabular}{cc}
\includegraphics[width=.5\textwidth]{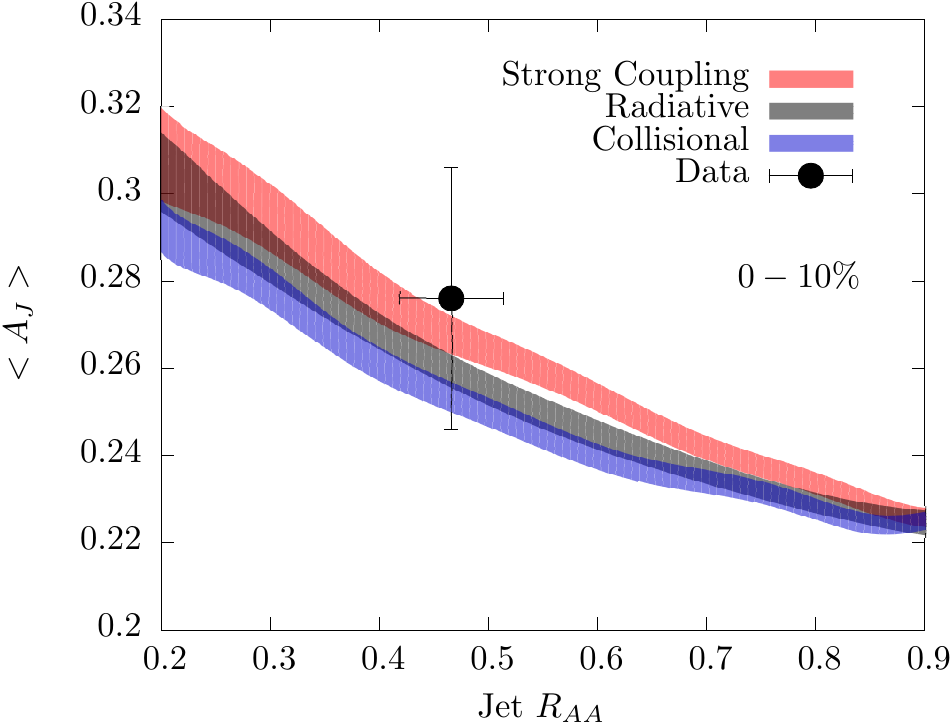}
&
\includegraphics[width=.5\textwidth]{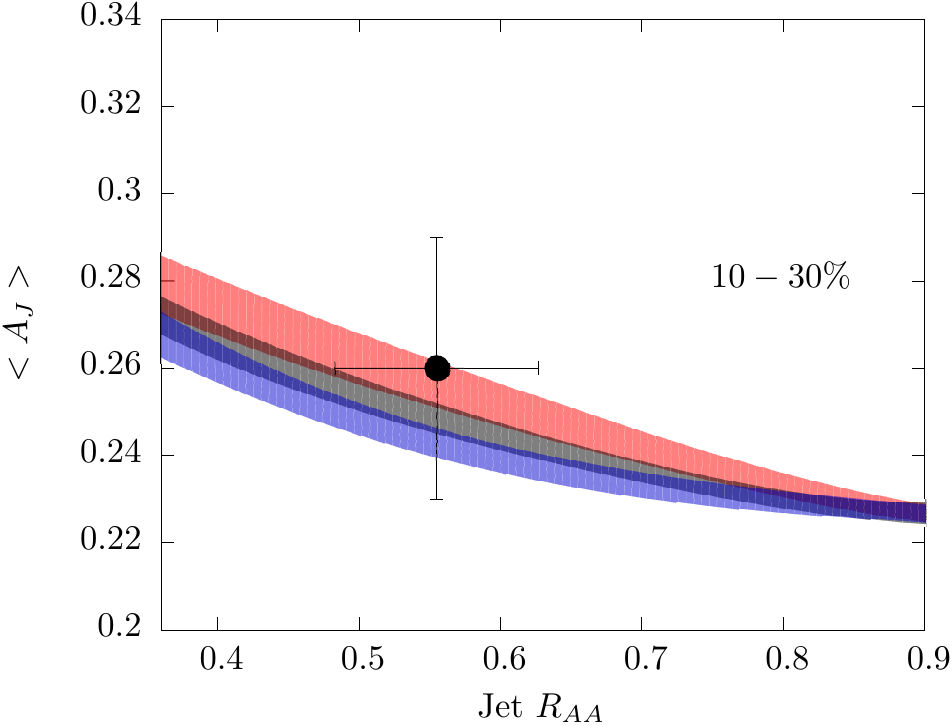}
\\
\includegraphics[width=.5\textwidth]{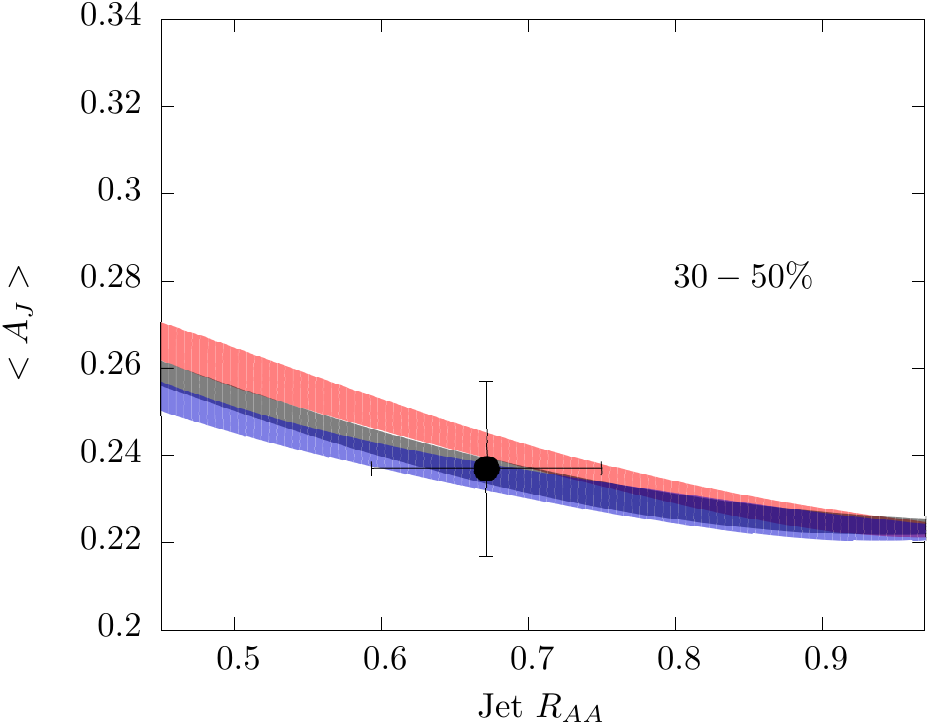}
&
\includegraphics[width=.5\textwidth]{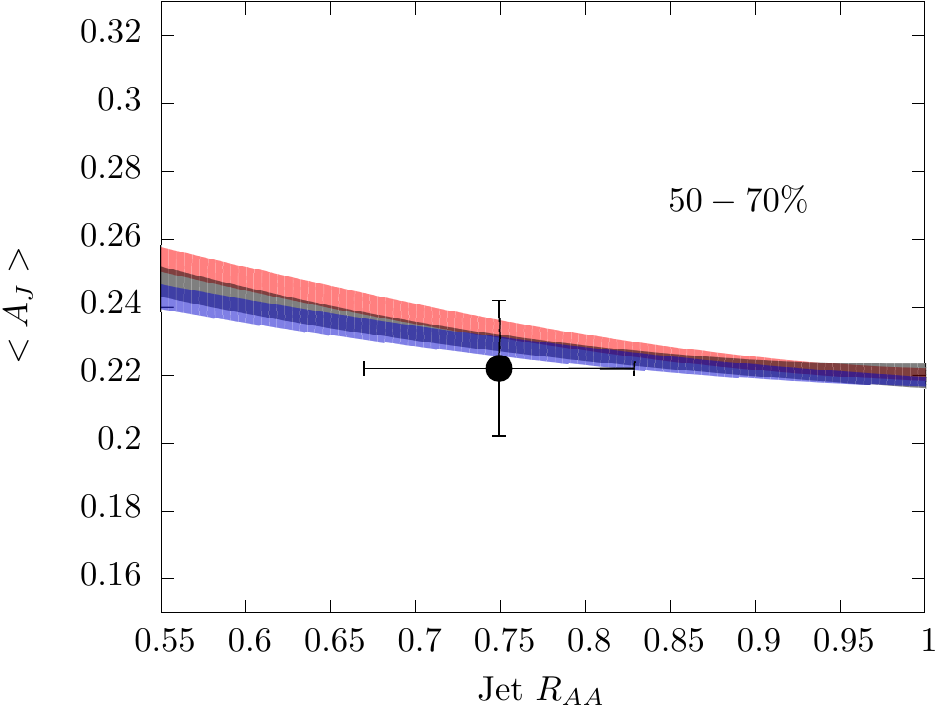}
\end{tabular}
\caption{\label{Fig:RAAvsAJ}  Mean di-jet imbalance as a function of jet suppression for three different models of jet energy loss. Data points are extracted from Refs.~\cite{Raajet:HIN} and \cite{Chatrchyan:2012nia}.}
\end{figure}

\subsection{Jet $\Raa$ and mean dijet asymmetry}

In order to get a better sense of the (in)sensitivity of the two classes of jet observables that
we have investigated so far to the mechanism by which the jet loses energy,
and in particular in order to look separately at the consequences of 
theoretical and experimental uncertainties, 
in Fig.~\ref{Fig:RAAvsAJ} we perform a 
parameter scan of the three models.
For each model, we perform a series of simulations with varying
values of the free parameter ($\aSC$, $\aR$ or $\aC$) in the 
expression for the energy loss rate $dE/dx$ 
(namely Eq.~(\ref{eq:Elrate}), (\ref{eq:Rad2}) or (\ref{eq:Coll}))
and compute both the mean asymmetry $\left< \AJ \right>$  
(for all dijet pairs with $\ponet> 120~{\rm GeV}$ 
and $\ptwot> 30~{\rm GeV}$)
and the jet suppression factor $\Raa$
(for jets with $100~{\rm GeV}<\pt<110~{\rm GeV}$) in a given centrality
bin. 
Each such scan over the value of the free parameter in one of the models yields a curve 
in the $(\left< \AJ\right>, \Raa)$ plane.  
For each model, we obtain a band in Fig.~\ref{Fig:RAAvsAJ}
that gives a sense of the theoretical uncertainty within
the given model by varying the critical temperature $T_c$ arising in
the hydrodynamic solution, as described in Section~\ref{MC}.
The different path length dependence  
of the three energy loss mechanisms is, in principle, 
reflected in the different shapes of the bands displayed in Fig.~\ref{Fig:RAAvsAJ}. 
These differ the most in the most central bin where, despite the width introduced by the theoretical uncertainty, the different behavior in the different models is distinguishable.  However, the differences between the
models are small compared to the present experimental uncertainties.
Consistent with what we have seen in Fig.~\ref{Fig:Asymcomp},
for $\Raa$ values close to the experimental ones the mean asymmetry of the three models is similar,
but with the strongly coupled model yielding slightly bigger asymmetries.  
For comparison, we have also plotted  the corresponding experimental data points which we extracted\footnote{
Since CMS uses different centrality bins for its $\left< A_J \right>$  and $\Raa$ measurements, we combined  the experimental values of $\Raa$ 
from CMS' $0-5\%$ and $5-10\%$ centrality bins, and their  measurements of $\langle A_J \rangle$
for their $10-20\%$ and $20-30\%$ bins.  In each such  combination, we weight the value of the observable
in each of the smaller centrality bins that we are combining by the ratio of  the number of jet events in that 
bin to the total number of jet events in the larger combined bin.
We extract these ratios 
from the forward calorimeter energy deposition distributions in 
jet triggered Pb-Pb events shown in Ref.~\cite{Chatrchyan:2012nia}.
}
from Refs.~\cite{Raajet:HIN} and \cite{Chatrchyan:2012nia}. 

The large systematic uncertainty in the determination of the mean ratio $\langle \ptwot/\ponet \rangle$ that
determines $\langle A_J\rangle$ is responsible  for the largest part of
the experimental error bars displayed in Fig.~\ref{Fig:RAAvsAJ}.  These large error bars, combined
with the smallness of the separation between the bands corresponding to the different
models, makes it impossible to use this analysis to  favor any of the models with any confidence.
However, the range of model parameters which can simultaneously accommodate the
measured values of the jet suppression and the dijet asymmetry is 
larger for the strongly coupled model.  
This corresponds to the slightly better fit to the dijet asymmetry 
data provided by the strongly coupled model
in Fig.~\ref{Fig:Asym}.
Perhaps the data therefore favor the strongly coupled model
very slightly.
At present, however, these data do not really discriminate among the models that we have explored,
given the current error bars and given the similarity between the predictions of these
three (very different) energy loss models  for the $\Raa$ and $A_J$ observables.
Although one could investigate whether the separation between the bands in Fig.~\ref{Fig:RAAvsAJ}
can be increased by using different ranges of $\pt$ in the evaluation of $\Raa$ or $\langle A_J\rangle$,
we do not anticipate reaching different conclusions until a time when the uncertainties in 
jet measurements at the LHC have been substantially reduced.

\begin{figure}
\centering 
\begin{tabular}{cc}
\includegraphics[width=.5\textwidth]{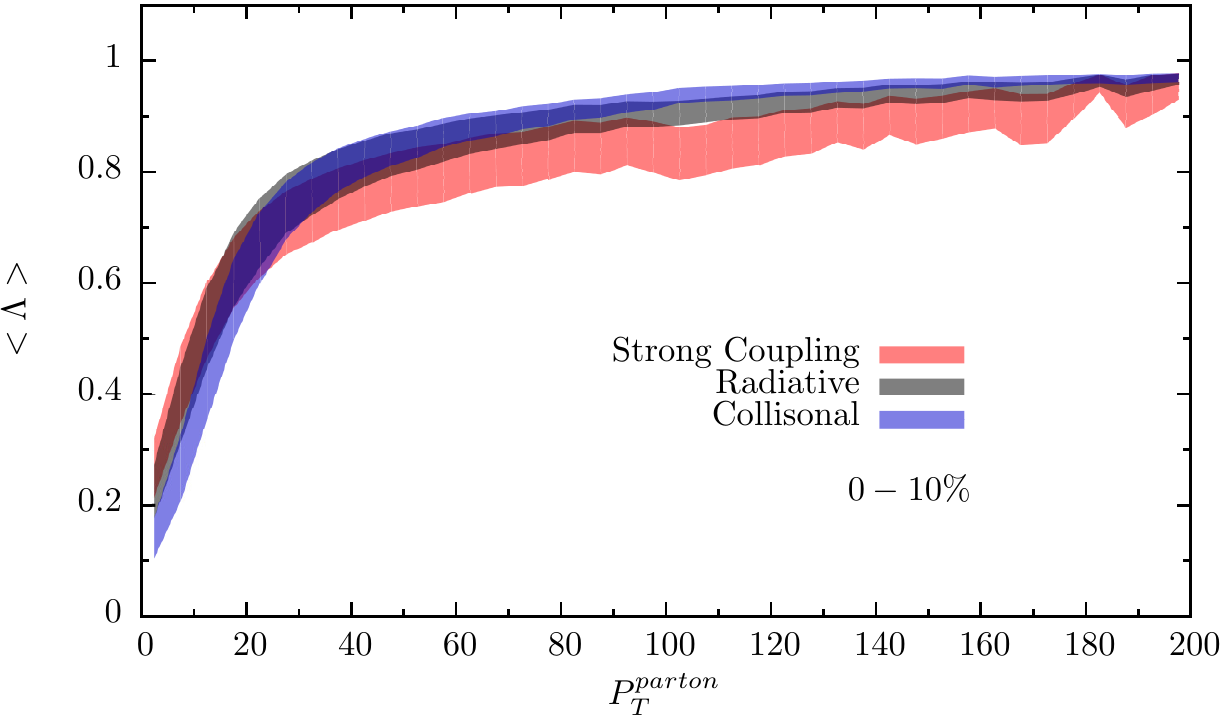}
&
\includegraphics[width=.5\textwidth]{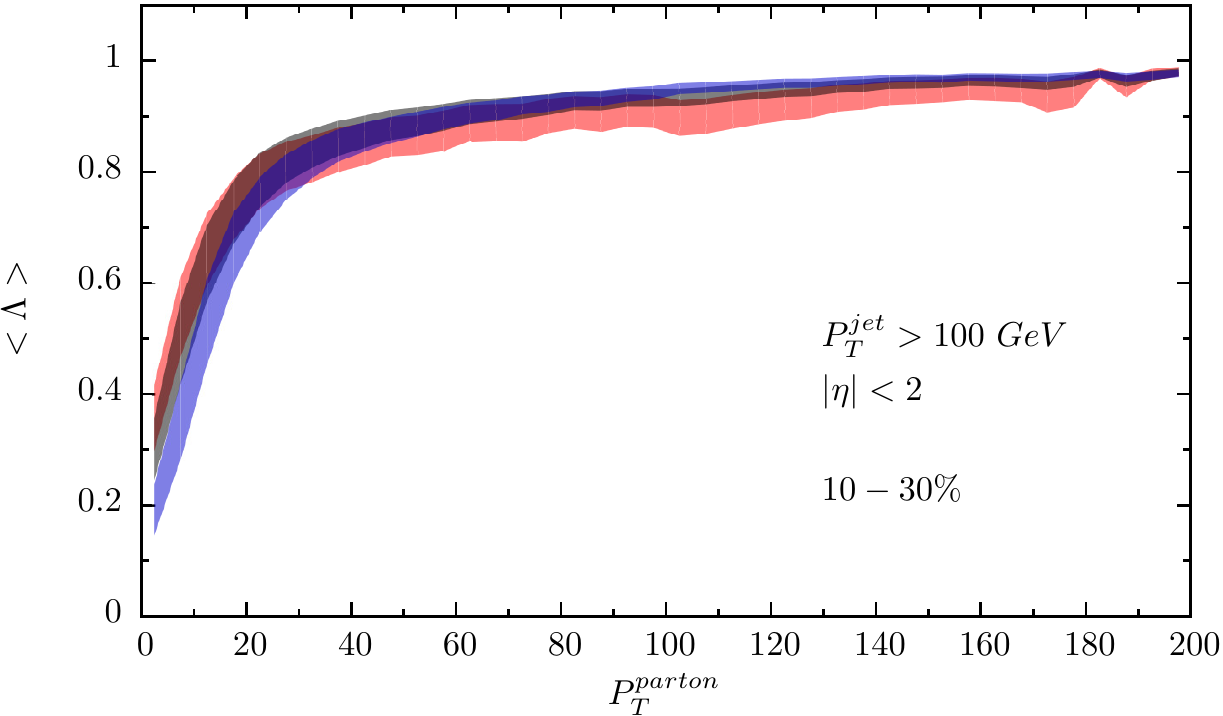}
\\
\includegraphics[width=.5\textwidth]{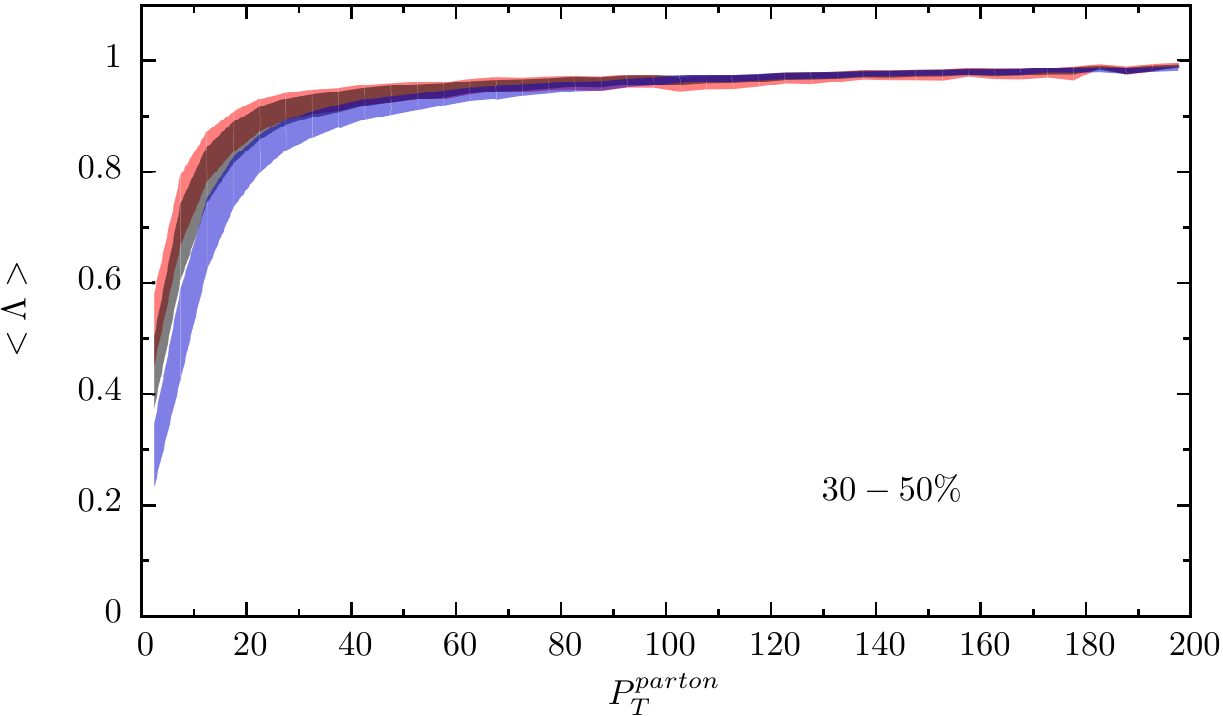}
&
\includegraphics[width=.5\textwidth]{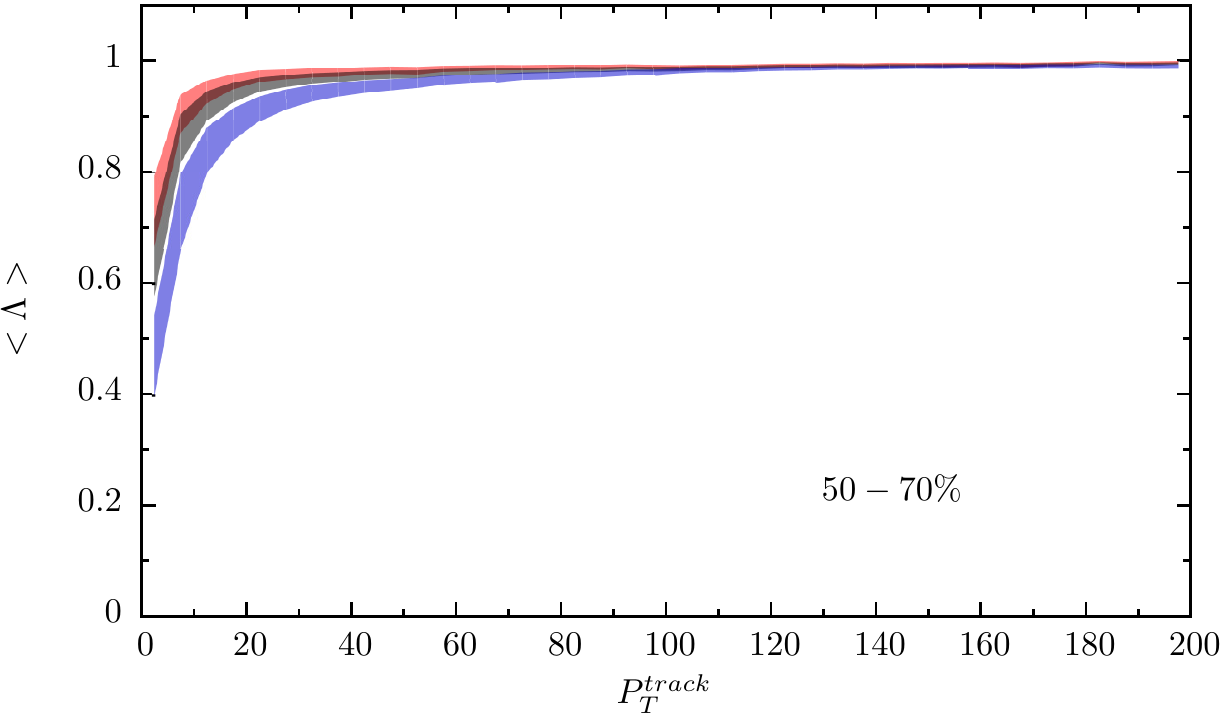}
\end{tabular}
\caption{\label{Fig:Lambdas}  Mean value
of the ratio $\Lambda$ between the transverse momentum $p_T^{\rm parton}$ 
of a parton in the reconstructed quenched jet to the transverse momentum that that
parton would have had in the absence of any medium. We plot $\langle \Lambda\rangle$
as a function of $p_T^{\rm parton}$ for jets with $\pt>100$~GeV in 
Pb-Pb collisions at the LHC in four centrality bins.  $\Lambda$ is not experimentally measurable; investigating
it nevertheless allows us to discern the effects of the differences between models.
 }
\end{figure}

\subsection{Energy lost by individual partons within a jet}

The insensitivity  of the inclusive jet observables
that we have looked at so far
 leads us  to consider more differential quantities. 
 We start 
 by studying the distribution of the
 energy lost by the individual partons within a reconstructed jet, although
 this distribution itself is not an experimental observable. 
 In Fig.~\ref{Fig:Lambdas}
we show the mean value
of the ratio $\Lambda$ of the transverse momentum of a parton after it has
been quenched by propagating through the plasma, 
 $p^{{\rm parton}}_T$, to the transverse momentum that that 
 parton would have had in the vacuum PYTHIA jet absent any quenching.
 We plot $\langle \Lambda \rangle$ as a function of the $\pt$ of the parton,
 averaged over all the partons in 
 the jets
 with total $\pt>100$~GeV in four different centrality bins. 
The average takes into account the fluctuation in $\Lambda$ 
induced both by variations in the path length through the medium
traversed by different jets 
as well as by the different pattern and times of branching that
can result in a parton in the final state jet with a given parton momentum. 
Although $\Lambda$ is not measurable, since knowing $\Lambda$ requires 
knowledge of the momentum 
that a parton would have had if there had been no medium present, it provides us with information
as to where differences among different models arise.

For all three models, the rough features of the distributions in Fig.~\ref{Fig:Lambdas} are qualitatively the same.
At high momentum, all the models feature a reasonably momentum independent $\left< \Lambda \right>$ which saturates at the highest momenta at roughly comparable values in all the models. 
This is a consequence of our fitting procedure: we have fixed the one free parameter
in each of the models so as to correctly describe $\Raa$ in a certain $\pt$ and centrality bin;
because of the steeply falling jet spectrum, $\Raa$ 
is only sensitive to whether jets that
start out with a given $\pt$ lose even a small amount of energy,
not to how 
much energy these jets lose on average and
not to the energy lost by jets that begin with higher $\pt$;
this in turn means that $\Raa$ is
most sensitive to the energy loss experienced by the hardest partons in a jet;
so, by fitting the parameter in each model to $\Raa$ 
we end up with the models having quite similar $\langle \Lambda \rangle$
at high parton momentum. 
Turning now to low parton momenta,
all three models also efficiently quench soft partons. 
For the collisional and radiative models,
this is a consequence of the fact that the energy loss rate $dE/dx$
in (\ref{eq:Rad2}) and (\ref{eq:Coll}) is independent of the energy
of the parton $E$.  This means that 
when we fix the parameter $\aR$ or $\aC$
by fitting to $\Raa$, ensuring some nonzero fractional energy loss for the highest momentum
partons, we end up with a larger
fractional energy loss for the lower momentum partons.
For the strongly coupled model, the quenching of soft partons
is enhanced by the Bragg-like behavior of the energy loss, with $dE/dx$ in (\ref{eq:Elrate})
rising rapidly as $x$ approaches $x_{\rm stop}$ and
the parton becomes soft.   

While the basic qualitative features of the
$\langle \Lambda \rangle$ versus parton $\pt$ curves plotted
in Fig.~\ref{Fig:Lambdas} are similar for all three models,
the quantitative shapes of the curves are
different for the different models.
The strongly coupled energy loss model yields a flatter curve
than do the other models, with a lower asymptotic value at high momentum and softer turn 
over for the most quenched partons than in the other two models. 
The collisional model, in which  $dE/dx$ has no path length dependence, has
the steepest behavior in Fig.~\ref{Fig:Lambdas}.
This correlation between the path length dependence of the energy loss model
and the behavior of the $\langle \Lambda \rangle$ curves in Fig.~\ref{Fig:Lambdas}
is easy to understand.  
Softer partons are in general created later and so travel
less distance in the plasma, meaning that if the rate of energy loss $dE/dx$
increases with distance traveled, as in the case of weakly coupled radiative
energy loss (\ref{eq:Rad2}) or the strongly coupled energy loss (\ref{eq:Elrate}),
the fractional energy lost by the soft partons is less than in the case of
collisional energy loss, where $dE/dx$ is independent of $x$.
So, the more pronounced the $x$-dependence of $dE/dx$ the flatter
the $\langle\Lambda\rangle$ vs. $\pt^{\rm parton}$ curve in Fig.~\ref{Fig:Lambdas}
should be, as indeed is seen in the Figure.
This more pronounced path length dependence is also responsible for the larger width of the theoretical 
uncertainties, since the relative enhancement of late time quenching makes the model more sensitive to the temperature at which quenching is turned off. 

\subsection{Fragmentation function ratio}
\label{sec:FFR}

\begin{figure}
\centering 
\begin{tabular}{cc}
\includegraphics[width=.5\textwidth]{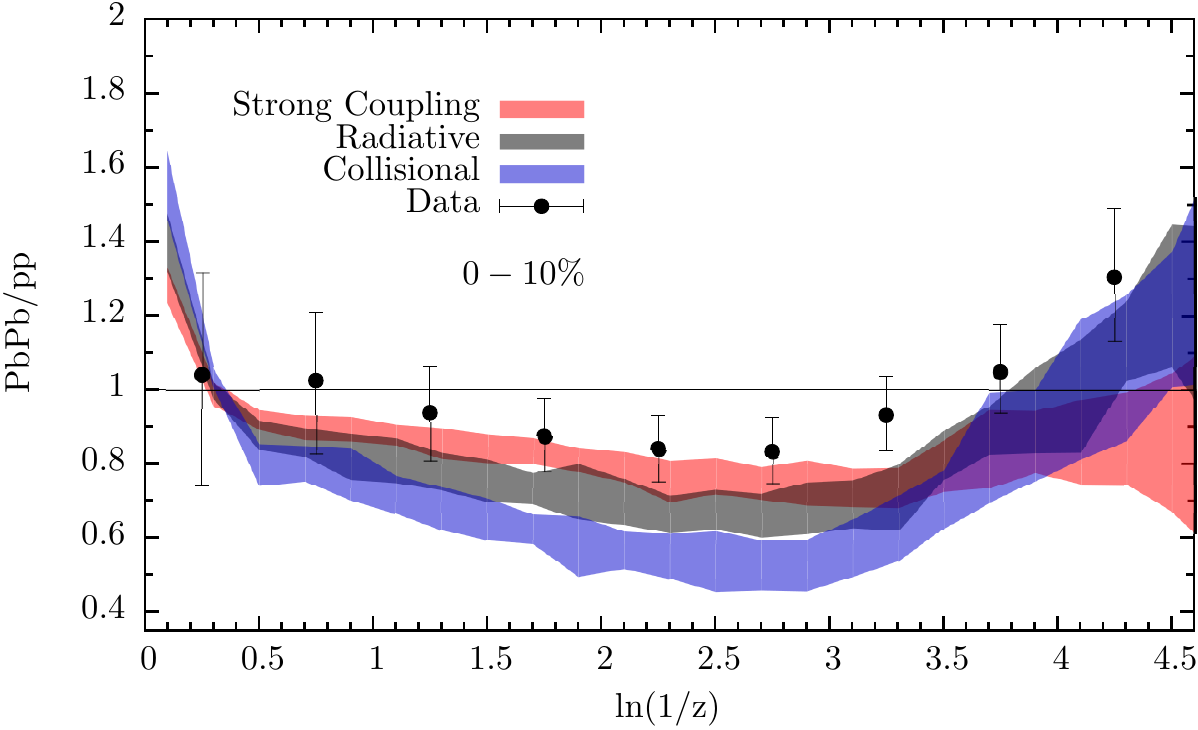}
&
\includegraphics[width=.5\textwidth]{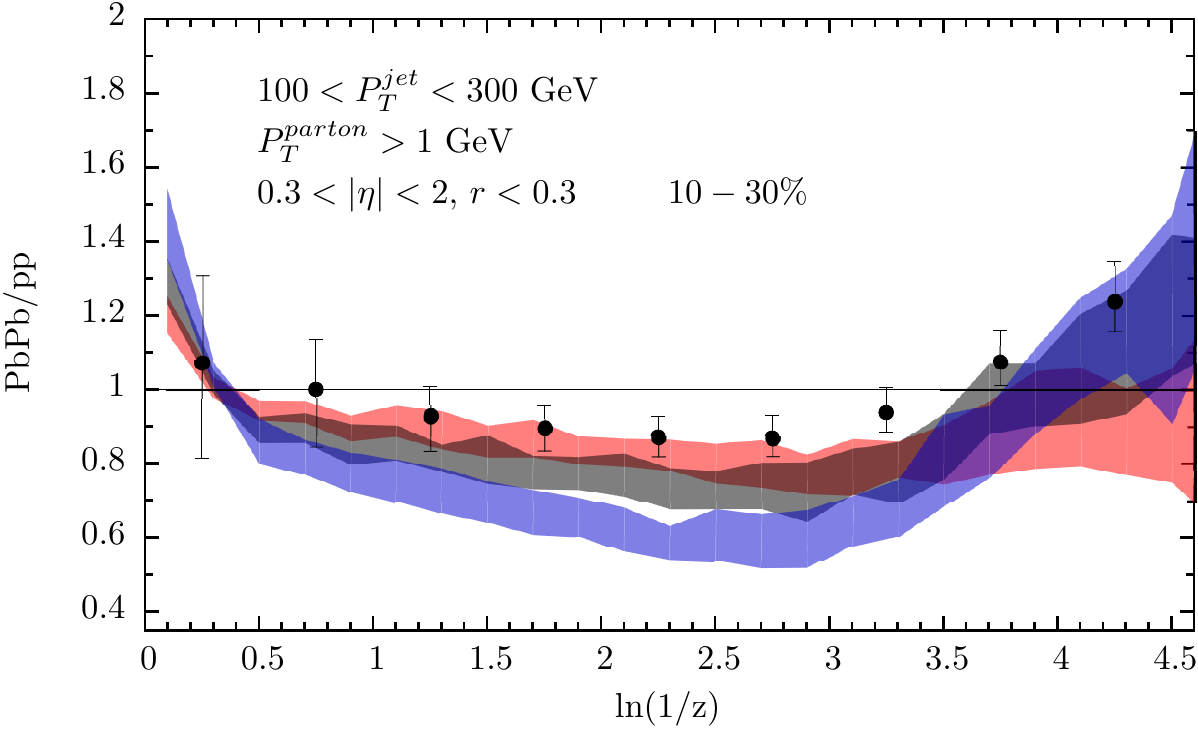}
\\
\includegraphics[width=.5\textwidth]{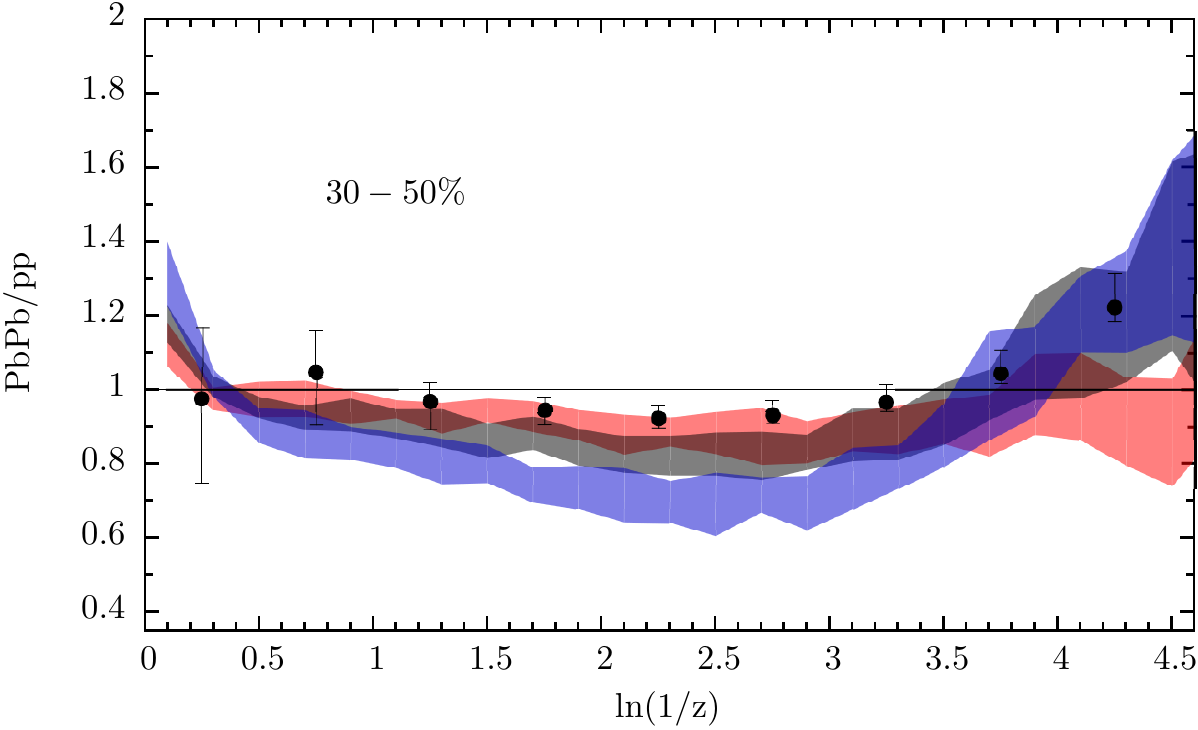}
&
\includegraphics[width=.5\textwidth]{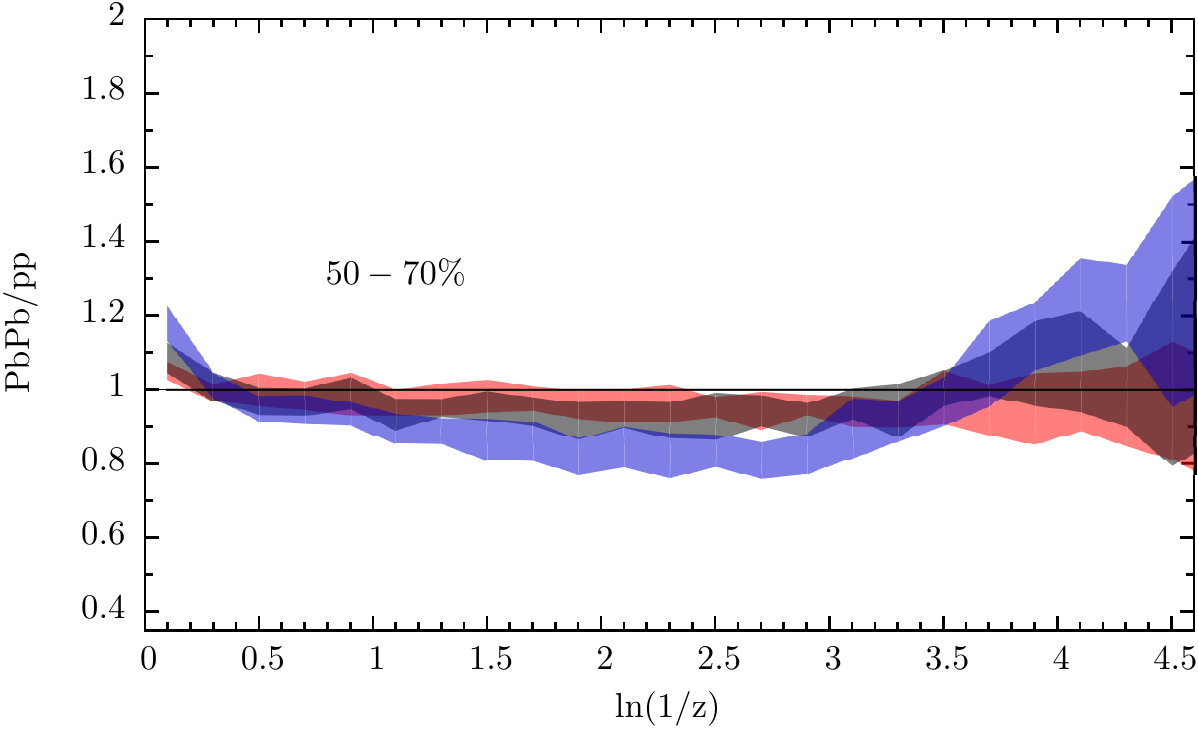}
\end{tabular}
\caption{\label{Fig:FFR}  Ratio of the partonic level fragmentation function for our
quenched jets to that for proton-proton jets 
in the same reconstructed jet $p_T$ interval $100< p_T < 300$~GeV, for jets 
with $0.3<\left|\eta\right|<2$ and for different centrality bins. The data points are the analogous experimentally measured fragmentation function ratio from Ref.~\cite{FFCMSt:HIN}, for hadrons rather than partons. 
}
\end{figure}

We have seen that inclusive jet observables
like the jet suppression factor $\Raa$ and the dijet
asymmetry $A_J$ are not particularly sensitive to the differences
between the three energy loss models that we are investigating.
We have also seen, however, that if we look at differences between the behavior
of partons within the jets with different $\pt$ we can find consequences of the 
different energy loss models.  This motivates us to investigate jet
fragmentation functions.  From our model, we can compute partonic fragmentation
functions, which are probability distributions for the fraction of the total jet momentum
that is carried by an individual parton in the final state jet.
In Fig.~\ref{Fig:FFR}, 
we show the ratio of the partonic fragmentation function for the
quenched jets in our hybrid strongly coupled model with $dE/dx$
as in (\ref{eq:Elrate}), as well as for the analogous weakly coupled radiative
and collisional energy loss models with
$dE/dx$ as in (\ref{eq:Rad2}) and (\ref{eq:Coll}), to the fragmentation function
for a PYTHIA jet in vacuum with the reconstructed jet energy in the same
interval as for the quenched jets.
This ratio is analogous to the fragmentation function ratio measured in Pb-Pb collisions at the LHC 
by both ATLAS and 
CMS~\cite{Chatrchyan:2012gw,FFCMSt:HIN,FFATLAS:HIN}, but 
of course they measure the fraction of the total jet momentum that
is carried by individual hadrons whereas our calculation is
performed at the parton level. 
As in the experimental analysis in Ref.~\cite{Chatrchyan:2012gw,FFCMSt:HIN}, we determine
the jet axis and 
momentum using the anti-$k_t$ algorithm with $R=0.3$ and we then
search for all the partons
in a cone with radius $r\equiv \sqrt{(\Delta \phi)^2 + (\Delta \eta)^2} <0.3$ 
centered on the jet axis
and use the projection of the parton momentum onto 
the jet axis to define $z=p^\mt{parton}_\parallel/p^\mt{jet}$. 
Since we have stopped the DGLAP evolution in PYTHIA for
each parton that reaches a minimum
virtuality
$Q_0=1$ GeV, we stop our computation of the fragmentation
function ratio in Fig.~\ref{Fig:FFR}  at 
$\ln(1/z)=\ln (100)= 4.6$. 
This also reduces our sensitivity to the effects of hadronization on 
the fragmentation function, which we are leaving out of our partonic
calculation.
For $z$ values smaller than our cut, the effects of hadronization become more important, since the dynamics 
of hadronization can soften particles below 1 GeV.

 The overall message from Fig.~\ref{Fig:FFR} is that 
 the fragmentation function of a quenched jet is very similar
 to that of a vacuum jet with the same energy as the quenched
 jet.  This was first pointed out in Ref.~\cite{Chatrchyan:2012gw} and remains the
 case in the data from Ref.~\cite{FFCMSt:HIN} that we have shown in the
 Figure, and it is also the case for our hybrid strongly coupled 
 model.  The collisional model that we have set up as a foil
 in this paper 
does not share this feature. The fragmentation function  ratio
predicted by the collisional model lies
 below the data over several decades of $1/z$, meaning
 that this model
can be  ruled out by the comparison of its fragmentation
 function to the data.
 The fragmentation function ratio predicted by our
hybrid strongly coupled model does best, comparing very well
with the data in Fig.~\ref{Fig:FFR}. The weakly coupled
radiative model fares in between, predicting
a fragmentation function ratio that is in some tension with the data,
particularly in mid-centrality collisions.  It should be noted that since in
the case of radiative energy loss the `lost' energy is expected to include
some moderate-$\pt$ particles that initially stay within the jet cone, 
the assumption that we are making in our implementation of this model
that all of the `lost' energy becomes soft particles moving in directions
that are uncorrelated with the jet direction may be suspect here. If 
so,
our calculation
of the fragmentation function in our implementation of radiative energy loss 
is incomplete.

Note that in  comparing our model calculations of the fragmentation function ratio to 
data in Fig.~\ref{Fig:FFR} we are ignoring
the softest part of the fragmentation function ratio shown
in the Figure. We do so for two reasons.  First, although we
have ended our partonic calculation at $Q_0=1$~GeV 
and cut the figure of at $z=1/100$ precisely
to reduce this problem, comparison of our partonic fragmentation functions 
to the data on hadronic
fragmentation functions may not be appropriate at the smallest $z$'s
we have plotted, given that hadronization tends to soften softer partons.
Second, the low $\pt$ particles that populate the smallest $z$
region that we have plotted in Fig.~\ref{Fig:FFR} have momenta
that are small enough that many of them could certainly come
from the thermal distribution of particles formed as the quark-gluon
plasma cools and hadronizes.   The background subtraction procedure
used in the analyses of  experimental data will subtract such particles,
on average, if they are uncorrelated with the jet direction.  This subtraction
may not be perfect, however, either because of fluctuations in the
bulk droplet of plasma or because
some of the energy lost by the jet, which we are assuming ends up 
as a little hotter or a little extra plasma, may also manifest itself in collective
motion of the plasma, meaning that although the `lost' energy becomes soft
particles these soft particles might not be completely uncorrelated with the
jet direction. For both these reasons,
the subtraction of whatever fraction of the `lost' energy
ends up in the jet cone may not be complete.
We have checked that adding only one soft particle per jet 
can result in a substantial upturn in the fragmentation function
ratio at $\ln(1/z)\gtrsim 4$, and for this reason we will not compare
to the data in this regime.

  The main features  of the fragmentation functions displayed  in Fig.~\ref{Fig:FFR} can be understood from the 
 distribution of quenching factors $\langle \Lambda \rangle$
  shown in Fig.~\ref{Fig:Lambdas}. 
  At $z\rightarrow 1$ the quenched (partonic) fragmentation function is close to the vacuum one, with 
  only a small enhancement observed. This enhancement 
  is a consequence of the depletion of soft fragments observed in all the models, which tends to make the 
  in-medium fragmentation functions harder than in vacuum. 
  While this enhancement is present in all three models, it is smallest 
  in the strongly coupled model, since the 
  quenching factor $\langle \Lambda \rangle$  is least dependent on the $\pt^{\rm parton}$
  in this model, see Fig.~\ref{Fig:FFR}.
  At intermediate $z$, all the in-medium fragmentation functions are depleted
  relative to the vacuum fragmentation functions. The $z$-values where such 
   depletion starts are correlated with the transverse momentum below which the quenching factor 
   $\langle\Lambda\rangle$ drops in Fig.~\ref{Fig:Lambdas}.  
   In the collisional and radiative models, this occurs at a higher momentum and, 
   as a consequence, the intermediate-$z$ depletion in the fragmentation function ratio is larger in
    these two models than for the strongly coupled model. The distinctions between the fragmentation function ratios of the three models
   at the lowest $z$'s plotted in Fig.~\ref{Fig:FFR} can also be understood in terms
   of features of Fig.~\ref{Fig:Lambdas}, but we have already explained why we will
   not focus on this region.

We have observed that the collisional model leads to a much stronger depletion of
the quenched fragmentation functions relative to what is measured
in data, over several decades of $z$.  This is a direct consequence of the 
lack of path-length dependence in $dE/dx$ in this model, meaning
that our conclusion that this model is disfavored seems robust.  
The radiative model seems to be marginally in agreement with the data.  
Remarkably,  the  weaker 
modification of the in-medium fragmentation function within the strongly coupled hybrid model  
achieves the best qualitative agreement with the fragmentation function ratio in the experimental data. 
A more quantitative, and more definitive, statement along these lines would require
including hadronization in our strongly coupled hybrid model, would
require investigating where the energy `lost' by the jet ends up rather than
just assuming that it becomes soft particles uncorrelated with the jet direction,
and would require including the soft particles corresponding to the plasma
itself in our model and subtracting them during jet reconstruction as in the
analyses of experimental data.  We leave all these investigations to future work.

\section{Conclusions, Discussion and a Look Ahead}
\label{sec:discussion}

\subsection{Conclusions}

We have seen in Section~\ref{cd} that our hybrid approach, with perturbative QCD (via PYTHIA)
describing the parton splitting that occurs within a jet while at the same time each parton in the
jet loses energy according to the expression  (\ref{eq:Elrate}) for 
$dE/dx$ for a light quark traveling through strongly
coupled plasma, derived via a holographic calculation in Ref.~\cite{Chesler:2014jva},
is very successful in describing the available jet data at the LHC.
After fixing the one free parameter in the model, defined in (\ref{eq:xstop}), 
using the measured value of the suppression factor $\Raa$ for jets in one $\pt$-bin in
the most central Pb-Pb collisions at the LHC, 
we obtain a completely satisfactory description of  the dependence of the jet $\Raa$ on both $\pt$ 
and centrality as well as of the dijet asymmetry $A_J$, including its centrality dependence.
In addition, we make predictions for the jet $\Raa$ at RHIC.
We also find that the (small) deviations between the fragmentation functions
of quenched jets measured in heavy ion collisions at the LHC
and those of vacuum jets with the same energy as the quenched
jets compare very well with the corresponding fragmentation
function ratios described by our hybrid model.

The above successes are important, but they 
should not be over-interpreted. 
The current uncertainties in the measurements of jet $\Raa$ 
translate into a significant dispersion in our theoretical computations, reflected
in the width of all the colored bands in our plots in Section~\ref{cd}.
And, partly as a consequence of these uncertainties and partly
as a consequence of the insensitivity of inclusive jet observables
to the mechanism by which energy is lost,
we have found that present measurements of the jet suppression factor $\Raa$ and the
dijet asymmetry $A_J$ are described almost as well if we use the models for $dE/dx$
motivated by weakly coupled radiative or collisional energy loss that
we have described in Section~\ref{elm}.
The comparisons between the partonic fragmentation function ratios that we can
compute in our models and the fragmentation functions measured at the LHC
that we have made in Section~\ref{sec:FFR} do favor the hybrid strongly coupled
approach over the model with collisional energy loss and, to some degree, over
the model with radiative energy loss.  However, this is a comparison between
a partonic calculation and a hadronic measurement, so perhaps we should
not take the fact that the data favors the strongly coupled energy loss rate
 as definitive.

The success (or partial success in the collisional case) of all these 
energy loss mechanisms, 
which arise from very different pictures of the underlying dynamics, crucially depends on the freedom to choose the overall strength of energy loss by fitting one model parameter to data. 
It therefore becomes important to confront the parameters extracted from data to
expectations from theoretical calculations.  We shall do this in Section~\ref{sec:parameters}.
We close in Sections~\ref{sec:weaknesses} and \ref{cfcadep}
with a look ahead in two senses, first with various ways that
our study could be improved and, second, with a suggestion for
an additional, more incisive, observable.

 \subsection{Significance of the extracted parameters}
 \label{sec:parameters}

 \begin{table}[t]
 \begin{center}
 \begin{tabular}{|c|c|c|c|}
 \hline
 & Strong Coupling & Radiative & Collisional \\
 \hline
 $ {\rm Parameter} $ & $ 0.26< \aSC < 0.35 $ & $ 0.81 < \aR < 1.60$  & $  2.5 <\aC < 4.2 $ \\
 \hline
 \end{tabular}
 \end{center}
 \caption{\label{alphatable} Values 
 of the fit parameters needed in the specification of $dE/dx$ in our three different energy loss models, in
 each case as extracted by comparing model predictions for $\Raa$ for jets with 100~GeV$<\pt <$110~GeV 
 in central
 Pb-Pb collisions at the LHC to experimental data.}
\end{table}

The three models for $dE/dx$ that we have tested in this paper each include
one free parameter that we have fitted to experimental measurements of
$\Raa$ for jets with 100~GeV$<\pt <$110~GeV
in central Pb-Pb collisions at the LHC.  We have collected the values of these parameters
obtained via fitting to this data in Table~\ref{alphatable}.
See Eqs.~(\ref{eq:xstop}),  (\ref{eq:Rad2}) and (\ref{eq:Coll}) for the definitions
of the parameters.

The values of $\aR$ and $\aC$ in the weakly coupled radiative and collisional models for $dE/dx$
obtained via our fit to data
should be compared to expectations based upon perturbative calculations given in
Eqs.~(\ref{eq:aRpert}) and (\ref{eq:aRpert2}) and in
Eqs.~(\ref{eq:aCpert}) and (\ref{eq:aCpert2}).
We see that our fit to data corresponds to a value of the strong coupling constant $\alpha_s$ that is 
smaller (larger) than the range $0.2<\alpha_s<0.3$ that we used
in making the estimate (\ref{eq:aRpert2}) for $\aR$ (the estimate (\ref{eq:aCpert2}) for $\aC$.)
In the case of radiative energy loss, as we discussed in Section~\ref{elm} it may be that
we are underestimating $\aR$ because we are neglecting the fact that much of the `lost' energy
is initially radiated in the form of gluons moving in the same direction as the jet,
meaning that some of this radiated energy may remain correlated with the jet direction.
If this is so, by neglecting this we would be overestimating the energy loss at
a given $\aR$ and hence our fit would be underestimating $\aR$.

As we have discussed in Section \ref{elm} and as is manifest in Eqs.~(\ref{eq:aRpert}) and (\ref{eq:aCpert}), 
because of rare radiative or collisional processes in which a large momentum is transferred
the perturbative evaluation of $\aR$ or $\aC$
leads to logarithms of ratios of scales, $B_{\rm rad}$ and $B_{\rm coll}$, 
which may depend on the kinematics of the colliding objects 
and whose evaluation is beyond the accuracy of current theoretical calculations. 
Since the precise expressions for both $B_{\rm rad}$ and $B_{\rm coll}$ are unknown,
it is best to think of our fits to data as constraining
the product of the appropriate power of $\alpha_s$ 
times the appropriate large logarithm, as in the middle expressions in 
Eqs.~(\ref{eq:aRpert2}) and (\ref{eq:aCpert2}).
Our fits yield relatively large values for this product, both in the case of radiative energy
loss and in the case of collisional energy loss. 
If the logarithmic corrections were small, as would be required for the simple perturbative expansion to be accurate,  our analysis would yield such large values of $\alpha_s$ that perturbation theory would clearly be invalid.
Or, if small values of $\alpha_s$ are chosen, as in the last expressions in
Eqs.~(\ref{eq:aRpert2}) and (\ref{eq:aCpert2}), then the logarithms become large which
again invalidates the simple perturbative expansion, in this case pointing towards the need
for a resummation
as discussed in Refs.~\cite{CasalderreySolana:2007sw,Iancu:2014kga,Blaizot:2014bha}.
Note also that despite our simplified approach to energy loss, our results are 
compatible with those of more sophisticated approaches, such as those described 
in Ref.~\cite{Burke:2013yra}, when the large logarithms are evaluated as prescribed in those works. 
The bottom line for the two weakly coupled models that we have introduced as benchmarks
is that within our model context they can describe LHC data on jet $\Raa$ and the dijet
asymmetry $A_J$ if we choose values of the single parameter in each model that correspond
to values of $\alpha_s$ that are large enough to make the reliability of a perturbative calculation 
questionable.  At the same time, as we saw in Section~\ref{sec:FFR} the 
collisional model cannot reproduce LHC data on the fragmentation function ratio
and the radiative model is in some tension with this data, at best in marginal agreement
with it.

We now turn to the strongly coupled model. The comparison of the value of 
$\aSC$ that we have obtained via fitting our results to jet observables
measured in heavy ion collisions at the LHC
to the value obtained in 
theoretical calculations performed holographically, {\it i.e.} via gauge/gravity duality,
is of necessity uncertain.  The holographic calculations that we have 
employed were done in large-$N_c$, strongly
coupled ${\cal N}=4$ SYM
theory, not in QCD. There are by now large classes of theories with known
gravitational duals, but the gravitational dual of QCD itself (if one exists) is not known.
Present holographic calculations are therefore best used to gain qualitative insights, like
for example the form of $dE/dx$ in (\ref{eq:Elrate}) and the parametric dependence
of $x_{\rm stop}$ in (\ref{eq:xstop}).  But there is no one right answer for
how to compare a numerical value of $\aSC$ extracted via comparison
to experimental measurements --- of course in QCD --- to a numerical
value of $\aSC$ computed in ${\cal N}=4$ SYM theory.   That said,
it is a generic expectation that the stopping distance $x_{\rm stop}$ 
will be longer, meaning that $\aSC$ will be smaller, 
in strongly coupled QCD plasma than in strongly
coupled ${\cal N}=4$ SYM plasma with the same temperature because
QCD has fewer degrees of freedom than ${\cal N}=4$ SYM theory by a factor
$\approx 0.4$.  There are various prescriptions in the literature for how this
reduction in the energy density of the plasma at a given temperature may
affect holographic calculations of various quantities, but this has not been investigated
for the stopping distance of a light quark.  And, of course, the QCD plasma differs from
that in ${\cal N}=4$ SYM theory in other ways also.

The comparison of the value of $\aSC$ that we have extracted via
comparison with data to theoretical expectations originating in holographic calculations
is further complicated by the fact that, as we have discussed in Section~\ref{elm},
theorists have developed several different
ways of modeling jets in ${\cal N}=4$ SYM theory, given that
jets are not actually produced in hard processes
in this theory.
Different values of $\aSC$ are obtained in ${\cal N}=4$ SYM theory depending
on whether a jet is modeled as a single string moving through the plasma,
in which case $\aSC^{{\cal N}=4}=1.05\lambda^{1/6}$~\cite{Chesler:2008uy}, or via
analyzing the decay of a virtual external $U(1)$ field into ${\cal N}=4$ SYM
matter with initial virtuality $q$ and initial position in the holographic
direction $D/q$ with $D$ an unknown factor that is of order
unity, in which case $\aSC^{{\cal N}=4}=1.24 D^{1/3}$~\cite{Arnold:2010ir}.
Although these two estimates
of $\aSC$ differ parametrically, the first being of order $\lambda^{1/6}$
while the second is of order unity,
their numerical values are similar. 
If we set $N_c=3$, the 't~Hooft 
coupling is $\lambda\equiv g^2 N_c = 12 \pi \alpha_s$ meaning that if 
we choose $0.2<\alpha_s<0.3$ this corresponds to $7.5 < \lambda <  11.3$
or $1.4< \lambda^{1/6} < 1.5$.  So, combining the two estimates, 
we learn that if we apply an ${\cal N}=4$ SYM theory calculation
done with $N_c\to\infty$ and $\lambda\to\infty$ to ${\cal N}=4$ SYM theory
with $N_c=3$ and $7.5 < \lambda < 11.3$ we conclude that $1.2 \lesssim \aSC^{{\cal N}=4} \lesssim 1.6$,
with the lower end of the range 
uncertain by a factor that is of order unity.
From this we conclude
that the value of $\aSC$ that we have extracted by comparing our results
to experimental data on $\Raa$ for jets  in the QCD plasma produced 
in LHC collisions is smaller than that in ${\cal N}=4$ SYM theory by
a factor of about 1/3 to 1/4, meaning that $x_{\rm stop}$ is longer
in the QCD plasma produced in a heavy ion collision than
in the ${\cal N}=4$ SYM plasma by a factor of about 3 to 4.

We conclude that the hybrid strongly coupled approach to jet quenching that
we have developed is in good agreement with all the various measured jet observables
to which we have compared it in Section~\ref{cd} when we take all the
parametric dependence of $dE/dx$ and $x_{\rm stop}$ from the 
expressions (\ref{eq:Elrate}) and (\ref{eq:xstop}) derived for the
${\cal N}=4$ SYM plasma, and set the  numerical value of $x_{\rm stop}$
in the QCD plasma longer than that in the ${\cal N}=4$ SYM plasma as expected,
longer by a factor of 3 to 4.

\subsection{Opportunities for improvements to our implementation}
\label{sec:weaknesses}

Although we have found that the inclusive jet observables $\Raa$ and $A_J$ have
limited discriminating power in differentiating between different energy
loss mechanisms, the success of the hybrid strongly coupled model
that we have developed in describing these data is encouraging.  The 
comparative success
of the hybrid strongly coupled model relative to the radiative model and, in particular, relative 
to the collisional model
in describing the data on fragmentation function ratios provides
 further encouragement.
What we have done is, however, only an initial exploratory study.  We are much more
confident in the value of our hybrid approach 
than in the specifics of the model implementation
that we have pursued in detail because we have made many simplifying
assumptions in  implementing our hybrid approach.
Here we summarize some of the main simplifications, all of which
represent opportunities for future improvements.  Such improvements
are well motivated indeed, given the increase in the 
quantity and quality of data on jet observables at both the LHC and RHIC anticipated
in the near future.

Some of the improvements that should be investigated come from the phenomenological
aspects of our model.  For example, our study should be repeated using solutions to three-dimensional
viscous hydrodynamics rather than the 
boost-invariant solution
to ideal hydrodynamics
that we have employed.  And, the effects of adding hadronization to the model should be studied,
as although this would open up new uncertainties it would also open up the possibility of
comparing to new observables.  As we have discussed in Section~\ref{sec:FFR}, it would
be of considerable interest to try to follow the energy lost by the quenched jet and to investigate
the degree to which the fraction of the `lost' energy that happens to become
soft particles within the jet cone
is or is not
subtracted during the jet reconstruction procedure used in the analysis of experimental data.

There are other improvements that should be investigated that reside within the holographic calculations
that yield results like (\ref{eq:Elrate}) that we have employed.  This list is fairly standard,
applying just as much here as in the many other contexts in which holographic calculations 
have been employed to gain qualitative insights into strongly coupled gauge theory
plasma and the dynamics of heavy ion collisions.  For example, one can ask about
finite $N_c$, finite $\lambda$, and nonzero $N_f/N_c$ corrections to (\ref{eq:Elrate}), or about
how this result changes in a strongly coupled theory that is not conformal.  

The opportunities for improvement that are more unique to the approach that we have introduced
in this paper reside in the hybridization of weakly coupled and strongly coupled dynamics
that is at the core of our approach.  To these we now turn.

A simple kinematic effect that we have neglected is 
the reduction in the phase space for the fragmentation of a parton
in the PYTHIA shower as a consequence of the energy loss that we have
added.
Although we have assumed that the energy loss results from processes
with small momentum transfer to or from the medium and therefore does
not modify the probabilities for the hard splitting processes, in reality
the reduction in the phase space for splitting will lead to some suppression
in the rate of splitting.  
While this effect is small for
the first energetic splittings, in the final stages of the shower
it may be more significant.
Given that $dE/dx$ in (\ref{eq:Elrate})
increases with increasing $x$, making all the partons live a little longer
will increase the effects of jet quenching if $\aSC$ is not modified which,
in isolation, 
would reduce the fitted value of $\aSC$.  At the same time, delaying
splitting will reduce the number of partons in the shower which could
reduce the effects of jet quenching for a give $\aSC$, resulting
in an increase in the fitted value of $\aSC$.

There is a second effect that works in the opposite direction
to the one above: as the partons in the shower interact
via multiple soft interactions with the medium these interactions may induce
additional splitting in the shower. Medium-induced splitting is of course at
the core of the weakly coupled radiative energy loss mechanism. 
Adding this physics would push in the opposite direction to
that above.   
It is hard
to see, however, how this could be done without paying the price of introducing
at least one further parameter that would have to be fit to data.  One of the
virtues of our present implementation is its minimalism.  This improvement, 
and many of the other improvements that we enumerate here, would 
reduce the minimalism of the approach.  As more data, more precise data
and data on more observables, becomes available this may become a price
worth paying.

A particularly important effect that we have not included in our computation is the kicks in
transverse momentum (transverse to the initial jet direction) that the fragments in the
shower will all pick up as they propagate through the medium, losing energy.
For simplicity, we have assumed that all the in-medium partons maintain their 
direction of propagation. The inclusion of transverse momentum 
broadening would have little effect on $\Raa$, which is dominated by the hardest fragments, and therefore
would not have much effect on the extracted value of $\aSC$.
However, as stressed in Ref.~\cite{CasalderreySolana:2010eh}, it would increase the dijet 
imbalance somewhat, since some of the soft fragments would get kicked out of the jet cone.  
We should mention, however, that this effect is unlikely to be pronounced
because partons in the shower
that become soft due to energy loss 
are very likely already being removed from the jet via
the consequent large Bragg-like increase in $dE/dx$ in (\ref{eq:Elrate}).
Including transverse momentum broadening would make it possible
to interpret other interesting observables.  For example, in our present
calculation our dijets are just as back-to-back as dijets in proton-proton collisions.
This is consistent with present data on the distribution of the azimuthal angle separating
jets in a dijet pair~\cite{Aad:2010bu,Chatrchyan:2011sx,Chatrchyan:2012nia}
and the distribution of the azimuthal angle separating the photon and the jet in 
gamma-jet events~\cite{Chatrchyan:2012gt}.  However, at present it 
would not be sensible for us to compare our model to these
data since there is no way within our model for these angular distributions to
be different in Pb-Pb collisions than in proton-proton collisions.
After adding transverse momentum broadening to our model, we could then
use the data that (at present) show no significant change in the
distribution of the dijet or photon-jet azimuthal separation angle from
proton-proton to Pb-Pb collisions to constrain the new component of the model.
We can further imagine using this data and a suitable variant of our hybrid model
to separately constrain the probability
that a hard parton is scattered by a large angle, thus looking
for evidence of the presence of point-like quark and gluon quasiparticles ~\cite{D'Eramo:2012jh}.
So, incorporating
transverse momentum broadening into our hybrid approach would result in a loss in minimalism
and an increase in the number of parameters that would need to be fitted to data but it
would mean that the model could be confronted with data on further observables, including
the distributions we have just mentioned or, for example, 
various measures of jet shapes.  We have made no attempt to analyze such
observables in the present paper since medium-induced modification of jet shapes has to
depend sensitively on
transverse momentum broadening.

In this paper we have considered each of the three different expressions for
the energy loss rate that we have investigated in isolation.  Adding
medium-induced splitting and transverse
momentum broadening, which are both characteristic of radiative energy loss,
to the hybrid strongly coupled model would be a step in the direction
of combining the mechanisms that in this paper we have treated separately.
After all, even when the typical interactions with the medium are soft and strongly coupled, with momentum
transfers of order the temperature, the partons in the
jet could have rare semi-hard interactions with constituents of the medium, inducing
both gluon radiation and scattering of the parton by 
a substantial angle~\cite{D'Eramo:2012jh}.  Looking for direct evidence of this in the data would be very
interesting since at present there is no direct evidence for the presence of the
weakly coupled point-like scatterers that, because QCD is
asymptotically free, must be seen if the strongly coupled
liquid quark-gluon plasma is probed at short enough distance scales.
It is therefore worth modeling and, ideally, separating the effects of strongly coupled
energy loss in conjunction with effects of occasional medium-induced gluon radiation and/or
hard scattering.
A further motivation for incorporating transverse
momentum broadening is that even if the physics is entirely strongly coupled, multiple soft interactions
add up to give nonzero transverse momentum broadening that can be
substantial in magnitude~\cite{Liu:2006ug,D'Eramo:2010ak}.

Another feature of the dynamics of energy loss 
that we have not implemented is the effects of finite resolution on the interaction
between the shower and the medium. In a finite medium, the 
separation of the jet fragments in the transverse direction in position space 
as they propagate through the plasma must be finite. As has been explicitly shown for radiative 
processes~\cite{CasalderreySolana:2012ef}, structures with a transverse size smaller than 
a given resolution scale must act coherently 
as seen by the medium. This reduces the effective number of propagating partons 
seen by the medium, and makes the `effective partons' harder than anticipated.
If $\aSC$ is left unchanged, these dynamics would tend to increase $\Raa$, 
reduce the dijet asymmetry and make the
fragmentation functions more similar to their vacuum counterparts.    Of course,
including these effects would result in a larger fitted value of $\aSC$.
At present no implementation of the effects of finite transverse resolution
is known at strong coupling, meaning that we have no evaluation of
the appropriate resolution scale for a strongly coupled plasma and meaning that
this investigation remains for the future.

Much remains to be done.  It will be interesting to see how robust the conclusions of our
study are as these further effects are included and as further observables become accessible
within our hybrid approach.

\subsection{\label{cfcadep}Distinctive species dependence and discriminating observables}

It is clearly important to find other less inclusive jet observables,
in addition to the fragmentation function ratios that we have analyzed, that can be measured
and that can further discriminate among different energy loss mechanisms.
There is one salient, and quite possibly very significant,
distinction between the models that we have introduced
that we have not utilized at all: the dependence
of the rate of energy loss $dE/dx$ on the color charge of 
the propagating hard parton.  
We have seen in Section~\ref{elm} that for both weakly coupled
energy loss mechanisms, namely radiative
energy loss as in (\ref{eq:Rad2})
and collisional energy loss as in (\ref{eq:Coll}), the ratio of
$dE/dx$ for quarks to that for gluons is $C_F/C_A=4/9$.
In contrast, in the strongly coupled calculation the stopping
distance (\ref{eq:xstop}) for quarks is longer than that for gluons
only by a factor of $(C_A/C_F)^{1/3}$.
This different color charge scaling means that even if parameters are chosen 
such that the overall magnitude of the energy loss is comparable in
the different models, in the strongly coupled model the amount
of energy lost by quarks and by gluons should be more similar
to each other while in the weakly coupled models they should differ more.
While the dependence of jet observables on 
this scaling is not straightforward to infer
because a jet that is initiated by a quark contains many gluons
in its fragments and vice versa,
the difference among models as to how $dE/dx$ depends on $C_F/C_A$
will leave an imprint in the suppression pattern of jets initiated by quarks as compared
to that of jets initiated by gluons.

\begin{figure}[tbp]
\centering 
\includegraphics[width=.7\textwidth]{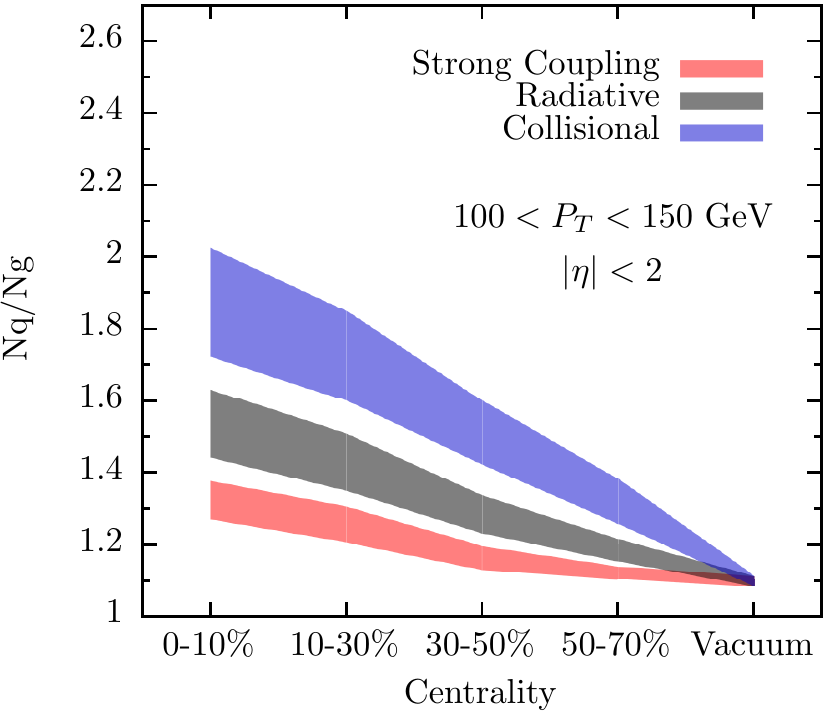}
\caption{\label{Charge_dep} Predictions of our model, with the three different
mechanisms for energy loss that we have investigated, for the ratio of the number of quark-initiated jets to 
the number of gluon-initiated jets as a function of centrality
for jets whose transverse momentum, after quenching, lies in the range $100~{\rm GeV}<\pt<150~{\rm GeV}$.
The centrality dependence of this ratio is significantly smaller for our hybrid strongly coupled model than
for either of the weakly coupled models.}
\end{figure}

In Fig.~\ref{Charge_dep} we show the ratio of the number of quark-initiated jets to the number
of gluon-initiated jets for jets with   $100< p^{\rm jet}_T<150$ GeV
and pseudo rapidity $\left| \eta\right| <2$ as a function of centrality.  The right-most point shows this ratio for vacuum jets. 
The centrality dependence of the 
ratio of the abundances of the two types of jets
is a clear manifestation of the different rates of energy
loss suffered by quark-initiated and gluon-initiated jets.
For the hybrid strongly coupled model, whose $dE/dx$ depends most
weakly on the parton's color charge, the ratio plotted in Fig.~\ref{Charge_dep}
is relatively close to its vacuum value 
for all the centrality bins, reflecting the fact that the energy loss experienced
by quarks and gluons is relatively similar in this model. 
In contrast, 
both the collisional and radiative models 
exhibit a much more pronounced centrality dependence in the ratio plotted in
Fig.~\ref{Charge_dep}.  In more central collisions in which jet quenching is
more significant overall, the gluon-initiated jets suffer more energy loss 
than the quark-initiated jets because $dE/dx$ scales with $C_A/C_F$
and so the ratio of quark-initiated jets to
gluon-initiated jets in a given $\pt$-range must increase.
The effect is greater in the collisional model than in the radiative model
because, as we saw in Fig.~\ref{Fig:Lambdas}, in the collisional model
the quenching of soft particles is particularly efficient and gluon jets
tend to have a softer fragmentation pattern than quark jets.

From this study 
we conclude that, if it were possible for experimentalists to identify
jets as quark-initiated or gluon-initiated, comparing the jet suppression factor $\Raa$ for
these two classes of jets would discriminate effectively between the three
different models of energy loss that we have considered.
Unfortunately, although there has been substantial
recent progress toward separating quark-initiated jets and gluon-initiated
jets in proton-proton collisions~\cite{Banfi:2006hf,Gallicchio:2011xq,Chatrchyan:2012sn,Krohn:2012fg,Gallicchio:2012ez,Larkoski:2013eya},
doing so in heavy ion collisions is sufficiently
challenging that it does not yet seem within reach.
One exception is jets produced back-to-back with a hard photon~\cite{Chatrchyan:2012gt}, since these
jets are predominantly quark-initiated jets.  Extending the implementation of our
hybrid approach to include gamma-jet events and using it to constrain the energy
loss of quark-initiated jets relative to that of all jets, and hence
to discriminate among models, will be of considerable interest.
Another exception
is b-tagged jets, a large fraction of which are b-quark-initiated jets.
Data on the suppression factor $\Raa$ for b-tagged jets with  transverse momenta
$\pt$ between 80 and 250 GeV~\cite{Chatrchyan:2013exa}
show no significant difference between their suppression and the suppression 
of inclusive jets in any of four centrality bins. Since at these very high values of $\pt$
the mass of the b-quarks should have little effect, b-quark-initiated jets in this regime
are a good proxy for quark-initiated jets, meaning that the data~\cite{Chatrchyan:2013exa}
favor energy loss models in which $dE/dx$ for a parton depends only weakly on the color charge
of that parton.  Although at present the experimental error bars are large,
these data already provide some further evidence
in support of the hybrid strongly coupled
model with $dE/dx$ as in (\ref{eq:Elrate}).
Reaching 
a firm conclusion also has to await further theoretical 
analysis of the energy loss of ultrarelativistic heavy quarks, and the jets initiated by them.
Holographic calculations
of the rate of energy loss of a heavy quark with mass $M$
moving slowly~\cite{CasalderreySolana:2006rq,Herzog:2006gh,Gubser:2006bz}
(with
a velocity such that its Lorentz boost $\gamma$ 
satisfies $\sqrt{\gamma}<M/(\sqrt{\lambda}T)$~\cite{Liu:2006he,Gubser:2006nz,CasalderreySolana:2007qw})
through strongly coupled plasma 
are well understood but the transition at larger $\gamma$ to the regime in which the
heavy quark behaves like a light parton is not yet understood.
Nevertheless, for b-quarks with $80~{\rm GeV}<\pt<250~{\rm GeV}$ it should be reasonable
to simply neglect the b-quark mass as we did above. Upon so doing,
we reach the conclusion that b-quark jets, and hence quark jets, are quenched
to the same degree as the mix of light-quark-initiated and gluon-initiated jets
found in inclusive jets.  This observation, together with the present data, 
favors the hybrid strongly coupled model
for $dE/dx$, as we have discussed. This suggests  that 
$\Raa$ for b-jets, $A_J$ for
dijets in which one or both of the jets are b-jets, and the b-jet fragmentation function
ratio can, if measured in heavy ion collisions and analyzed via our hybrid approach, 
yield
observables that discriminate effectively between energy loss models.
With further development, and in particular
with the investigation of these and other observables, the observation that at strong coupling
energy loss depends more weakly on the type
of parton (quark vs. gluon) may yield a robust signal for the strongly (or weakly) coupled
nature of medium-induced jet quenching.

Much remains to be done and many further observables
remain to be investigated. Further exploration of the hybrid approach that we have introduced in
this paper and its implementation via the strongly coupled energy
loss rate (\ref{eq:Elrate}) and the stopping distance (\ref{eq:xstop})
is strongly motivated
given how well the results we have obtained
agree with data on jet $\Raa$, the dijet asymmetry $A_J$ and,
to this point most discriminatingly, the fragmentation function ratio.
The hybrid approach has already provided us with a calculational framework
within which we can test strongly coupled predictions for jet quenching by
confronting them quantitatively with experimental measurements of jet
observables.  This demonstrates that this approach can now be used to
explore and subsequently test new observables.  Having the means to
quantitatively confront new ideas, like for example the relationship between
the centrality dependence of the ratio of the number of quark-initiated jets
and the number of gluon-initiated jets that remain in the final state, new
observables, and new data is critical if we are eventually to understand
the properties of the strongly coupled liquid quark-gluon plasma that
Nature has served us.

\vspace{0.5cm}
\noindent 
{\bf Note Added:} After the publication of this work we have found a small mistake in our implementation of the Glauber Monte-Carlo for the collision geometry, which resulted in an incorrect distribution of the points in the transverse plane at which our jets were created,  for collisions within each centrality bin.
After correcting this, 
the fitted values of $\aSC$, $\aR$ and $\aC$  in  the current version of this paper are slightly smaller than previously reported. 
This correction also results in small (hardly visible) changes to all the plots in this paper, which we have corrected. 
Neither the changes to the fitted values of the $\kappa$'s nor the changes to the figures affect any discussion or any conclusions of this work; we have made
no changes to the wording in any discussion of any figure or result. 
We have also reduced the $\pt$-cut of our  PYTHIA simulations from $70$~GeV to $50$~GeV to further reduce the sensitivity
of our computations to this cut.

\acknowledgments

We are grateful to Paul Chesler, Peter Jacobs, Andreas Karch, Yen-Jie Lee, Al Mueller, Gunther Roland, 
Jesse Thaler, Xin-Nian Wang and Korinna Zapp for  helpful conversations over the course of this work.
KR is grateful to the CERN Theory Division for hospitality
at the time this research was begun.
The work of JCS was  supported by a Ram\'on~y~Cajal fellowship.  The work of JCS and DP was 
supported by  the  Marie Curie Career Integration Grant FP7-PEOPLE-2012-GIG-333786, by grants FPA2010-20807 and  FPA2013-40360-ERC from the 
Ministerio de Econom\'ia y Competitividad, 
Spain,  by grant 2009SGR502 from the 
Generalitat de Catalunya 
and by the Consolider CPAN project. 
The work of DCG and KR was supported by the U.S. Department of Energy
under cooperative research agreement DE-FG0205ER41360.
The work of JGM was supported by Funda\c{c}\~{a}o para a Ci\^{e}ncia e a Tecnologia (Portugal) under  project  CERN/FP/123596/2011 and contract `Investigador FCT -- Development Grant'.


\end{document}